\documentclass[a4paper,12pt]{article}
\usepackage{epsfig}

\begin{document}

\def\beq{\begin{equation}}
\def\eeq{\end{equation}}
\def\bea{\begin{eqnarray}}
\def\eea{\end{eqnarray}}
\def\Dss{\frac{1}{3 s^2}}
\def\Dtt{\frac{1}{t^2}}
\def\Duu{\frac{1}{u^2}}
\def\Dtu{\frac{1}{t\,u}}
\def\Dts{\frac{1}{t\,s}}
\def\Dt{\frac{1}{t}}
\def\Du{\frac{1}{u}}
\def\Ds{\frac{1}{s}}
\newcommand{\dedouble}{ \stackrel{ \leftrightarrow }{ \partial } }
\newcommand{\deR}{ \stackrel{ \rightarrow }{ \partial } }
\newcommand{\deL}{ \stackrel{ \leftarrow }{ \partial } }
\def\dofig#1#2{\epsfxsize=#1\centerline{\epsfig{file=#2, width=14cm, 
height=5cm, angle=0}}}
\def\dofigs2#1#2#3{\centerline{\epsfxsize=#1\epsfig{file=#2, width=7.5cm, 
height=7.5cm, angle=-90}
\hfil\epsfxsize=#1\epsfig{file=#3,  width=7.5cm, height=7.5cm, angle=-90}}}
\def\makefigs4#1#2#3#4#5{
\centerline{\epsfxsize=#1\epsfig{file=#2, width=7.5cm, 
height=7.5cm, angle=-90}
\hfil\epsfxsize=#1\epsfig{file=#3,  width=7.5cm, height=7.5cm, angle=-90}}
\centerline{\epsfxsize=#1\epsfig{file=#4, width=7.5cm, 
height=7.5cm, angle=-90}
\hfil\epsfxsize=#1\epsfig{file=#5,  width=7.5cm, height=7.5cm, angle=-90}}
}
\renewcommand{\thefootnote}{\fnsymbol{footnote}}
\rightline{CERN-TH/2002-034} \rightline{HIP-2002-06/TH} 
\vspace{.3cm} 
{\Large
\begin{center}
{\bf Gravitational decays of heavy particles \\ in
large extra dimensions}

\end{center}}
\vspace{.3cm}

\begin{center}
E. Gabrielli$^{1,2}$ and B. Mele$^{3}$ \\
\vspace{.3cm}
$^1$\emph{ Theory Division, CERN, CH-1211 Geneva 23, Switzerland}
\\
$^2$\emph{
Helsinki Institute of Physics,
     POB 64,00014 University of Helsinki, Finland}
\\
$^3$\emph{Istituto Nazionale di Fisica Nucleare, Sezione di Roma
and Dip. di Fisica, Universit\`a La Sapienza,
P.le A. Moro 2, I-00185 Rome, Italy}
\end{center}

\vspace{.3cm}
\hrule \vskip 0.3cm
\begin{center}
\small{\bf Abstract}\\[3mm]
\end{center}
In the framework of quantum 
gravity propagating in large extra dimensions, we analyze
the inclusive radiative 
emission of Kaluza-Klein spin-2 gravitons in the two-fermions decays of
massive gauge bosons, heavy quarks,   
Higgs bosons, and in the two-massive gauge bosons decay of Higgs bosons.
We provide analytical expressions for the square modulus of amplitudes 
summed over 
polarizations, and numerical results for the widths and branching ratios.
The corresponding decays in the $Z$, top quark, and Higgs boson sectors 
of the standard model are analyzed
in the light of present and future experiments.

\begin{minipage}[h]{14.0cm}
\end{minipage}
\vskip 0.3cm \hrule \vskip 0.5cm
\section{Introduction}
After the recent proposal of Arkani--Hamed, Dimopoulos, and Dvali 
(ADD) on quantum gravity propagating in large extra dimensions \cite{ADD},
there has been an intense theoretical activity on this subject
\cite{GRW}-\cite{AS}.
In \cite{ADD}, it was pointed out that if compactified
extra dimensions exist, with only gravity propagating in
the bulk and standard matter with gauge fields
confined in the usual 3+1 dimensional space,
then the fundamental scale of quantum gravity could be much lower 
than the Planck scale $M_P$.
In particular the weakness of gravity might be due to the large size of the
compactified extra dimensional space.\footnote{ For a realization of 
large extra dimension scenarios in the framework of string theories, 
see \cite{ED_string}.}
Indeed, in this scenario, the Newton constant
$G_N$ in the 3+1 dimensional space is related to  the corresponding Planck
scale $M_D$ in the $D=4+\delta$ dimensional space, by
\beq
G_N^{-1}= 8 \pi R^{\delta} M_D^{2+\delta}
\label{newton}
\eeq
where $R$ is the radius of the compact manifold
assumed here to be on a torus.

Large extra dimensions can therefore provide a new
solution to the hierarchy problem and open new attractive scenarios
\cite{ADD}.
In particular, if $M_D\sim $ TeV then
deviations from the Newton law are expected
at distances of order $R < 10^{32/\delta -19}$ meters \cite{Gravity_test}.
The present experimental sensitivity in gravity tests
is above the millimeter scale, 
and the solution to Eq.(\ref{newton}),
with $M_D \sim $ TeV, requires $\delta \ge 2$.
A dramatic phenomenological consequence of this theory is
that  quantum gravity effects
could be sizeable already at the TeV scale, and could be tested at
present and future collider experiments.

After integrating out compact extra dimensions,
the Einstein equations in the four-dimensional space 
describe massive Kaluza-Klein (KK) excitations
of the standard graviton field.\footnote{
For a detailed discussion about the
effective four-dimensional theory, see  \cite{GRW}.}
These KK excitations
are very narrowly spaced in comparison to the $M_D$ scale,
the mass splitting ($\Delta m_G$) being of order
$\Delta m_G \sim \frac{1}{R}= 
M_D \left(\frac{M_D}{\bar{M}_P}\right)^{2/\delta}$, where
the reduced Plank mass $\bar{M}_P$ is defined as 
$\bar{M}_P^2=(8\pi G_N)^{-1}$.
The couplings of the KK gravitons to the standard matter and gauge fields
is therefore universal and equal to their zero modes, and hence
suppressed by $1/\bar{M}_P$.
On the other hand, in the case of inclusive production (or virtual exchange) 
of KK gravitons, remarkably, the sum over the allowed tower 
of KK states (which could be approximated by a continuos) gives a 
very large number. This number exactly cancels the suppression factor 
$\frac{1}{\bar{M}_P^2}$ 
associated to a single graviton production, replacing it 
by $\left(\frac{E}{\bar{M}_D}\right)^{2+\delta}$,
where $E$ is the typical energy of the process. 
Therefore, if $M_D$ is in the TeV range, quantum gravity effects 
might become accessible at future collider experiments.

Another interesting possibility for the solution of the hierarchy problem, 
as suggested by Randall and Sundrum \cite{RS}, is to have a non-factorizable 
geometry where the 4-dimensional massless graviton 
field is localized away from the brane where 
standard matter and gauge fields live.
The main signature of this scenario is quite
different from the one arising from the ADD scenario 
 in collider experiments \cite{RS_phen}.
Indeed, 
widely separated and narrow spin-2 graviton modes are expected.

In the present paper, we restrict our analysis to the ADD scenario.
In this framework, the relevant physical processes
in $e^+ e^-$ and hadron collider experiments have been first analyzed in 
\cite{GRW} (see also  \cite{PHEN_ED},\cite{higgs}).
They can be classified in: a) direct production of KK gravitons and 
b) virtual gravitons exchange. In the first case, the best 
signatures corresponding to the final state would be a photon associated 
with missing energy (in electron colliders) or jet + missing energy
(in hadron colliders). In the latter case, the
gravitons exchange will induce local dimension-eight operator (associated 
with the square of the energy momentum tensor)  that will affect the
standard four fermion interactions processes.
The main conclusion of \cite{GRW} is that searches at LEP2 and Tevatron 
can probe the fundamental $M_D$ scale up to approximately 1 TeV, 
while the CERN Large Hadron Collider (LHC) and  linear $e^+e^-$ colliders
will be able to perturbatively probe this scale up to several TeV's.

In the present scenario, for any new heavy particle 
with mass close to $M_D \sim $TeV, 
the gravitational radiation induced in its decays 
might become important.\footnote{This does not include
scenarios with brane deformations, see for instance 
Refs.\cite{fatbrane,rizzo}, where $KK(n)$ tower states of SM particles 
can have tree-level 2-body decays in $KK(n) \to KK(n-1) + G$.} 
Indeed, the suppression factor in the branching ratio 
will be given in this case by
$\left(\frac{M}{\bar{M}_D}\right)^{2+\delta}$, where
$M$ is the mass of the decaying state \cite{ADD}.
For instance, 
new particles at the TeV scale are  expected in some models
where the grand-unification scale is lowered down to the TeV scale by 
the appearance of new compact extra dimensions where 
standard model (SM) 
fields live 
\cite{TH_SM_ED}-\cite{PHEN_SM_KK}.
Such extra dimensions are a natural consequence of string theories with
large radius compactification. These scenarios could provide both
a natural explanation for the fermion mass hierarchy (since the
fermion masses evolve with the mass scale by a power law dependence
\cite{DDG1}), and a natural  higher-dimensional seesaw
mechanism for giving masses to light neutrinos\cite{DDG2}.  
Therefore, in a unified picture of gauge and gravitational 
interactions with unification
occurring at around the TeV scale, we should expect new Kaluza-Klein
excitations of the SM particles  at the TeV scale.
In the decays of such states, the gravitational radiation
could give rise to relevant effects.

The aim of the present work is to provide, in the framework of 
gravity propagating in large extra dimensions,
analytical and numerical results for both differential widths and 
inclusive branching ratios of gravitational decays 
of heavy particles,  versus the number of extra dimensions.
In particular, we consider the following classes of decays
\beq 
V,\, H \to \bar{f_i}\,\,  f_{i,j} + G,~~~~~f_i \to f_j\,\,  V + G, ~~~~~
H \to V\, V  + G
\eeq 
where $V,~H$, and $f_{i}$ 
represent a generic massive gauge boson, Higgs boson, 
and fermion field (with $i \neq j$), respectively.
We retain the masses of all the particles in the final states,
except in the decay $f_i \to f_j\,\,  V + G$ where the
final fermion is assumed massless.
We then apply our results to the analysis  
of the $Z$, $W$, top quark, and Higgs boson decays in the SM.
On the other hand, our results can be easily applied to more general cases, 
too.

We restrict our discussion to the spin-2 gravitons, and
do not include 
the corresponding decay modes into scalar gravitons (graviscalars, with
J=0) since their amplitudes get smaller with respect to 
the J=2 ones, being suppressed by a term 
proportional to {\small $\omega=1/\sqrt{3(\delta+2)/2}$} \cite{HLZ}.
A special case is provided by  scenarios where the Higgs
boson can have a mixing to graviscalar field through the coupling to the 
Ricci scalar \cite{higgs},\cite{AS}. In these scenarios, 
the inclusive decay of the Higgs boson 
in all the allowed tower of KK graviscalars is very large, and 
leads in practice to a sizeable invisible width for the Higgs 
boson \cite{higgs}.

In the framework of gravity propagating in large extra dimensions, 
in \cite{Zpole} 
the decay $Z \to \bar{f}\, f + G$ 
has been analyzed for both J=2 and J=0 gravitons,
versus high precision LEP1 $Z$-pole data. 
Only numerical results are  provided for J=2.
As shown in section 3, our  results 
for the inclusive total width $\Gamma(Z \to \bar{f}\, f + G)$  
agree with \cite{Zpole}.

The paper is organized as follows. In section 2, we define
the interacting lagrangian describing massive gauge bosons coupled to 
fermions and Higgs fields, and the corresponding energy momentum
tensor which enters the coupling to the graviton field.
In section 3,4, and 5, we give the analytical and numerical results 
for widths and branching ratios, and discuss the corresponding 
decays for  the SM  $Z$/$W$, top quark, and Higgs boson, respectively.
In section 6, we present our conclusions. In appendix A1, we
report the relevant Feynman rules for the gravitational interaction vertices,
and in appendix A2 we give the analytical expressions for
the square modulus of the amplitudes.

\section{Effective Lagrangian}
The coupling of gravity with 
standard matter and gauge fields in D-dimensional space is given 
by the lagrangian ${\cal L}_D$ \cite{GRW}
\beq
{\cal L}_D = \frac{1}{\bar{M}_D^{1+\delta/2}}
\, T_{A B} \, h^{A B},~~~~~A=(\mu,i), 
~~\mu=0,\dots ,3, ~~~~i=4,\dots , D-1
\eeq
where $\bar{M}_D^{2+\delta}=\bar{M}_P^2(2\pi R)^{-\delta}$,
$T_{A B}$ is the energy momentum tensor, $h^{A B}$
the graviton field in a D-dimensional space, 
and the $A$ and $B$ indices refer to the D-dimensional space.
The sector of the energy-momentum tensor 
$T_{A B}$ containing standard matter and gauge fields 
is assumed here to have non-zero
component only along the $A,B= \mu, \nu$ directions.\footnote{ This can be realized
assuming that SM particles correspond to brane excitations
and the brane itself does not oscillate in the extra dimensions.}
After integrating out the compactified extra dimensions in the 
D-dimensional action, the resulting (effective) four dimensional 
theory is described by KK graviton fields $h^{(n)~\mu \nu}$
which have the same universal coupling
to the SM particles as their massless zero-mode (n=0) \cite{GRW}.
Then, in four dimensional space, the effective Lagrangian is given by
\beq
{\cal L}_{eff} = \frac{1}{\bar{M}_P}\sum_n
\, T_{\mu \nu} \, h^{(n)~\mu \nu}\, ,
\label{gravity}
\eeq
where $\bar{M}_P$ is the reduced Planck mass, and
$T_{\mu \nu}$ is the SM energy momentum tensor.

In this section, we fix our conventions for the Lagrangian 
$\cal{L}$ and its energy momentum tensor $T_{\mu\nu}$ 
which are relevant for the processes we are considering.
In particular, we generalize the  fermion fields ($f_i$) couplings 
of the SM in the weak gauge boson ($V$) 
and Higgs ($H$) sectors.
This parametrization might be particularly useful in a generalization
of the SM interactions including KK excitations of 
SM fields, when SM fields are assumed to propagate in other
extra-dimensions.

In Minkowski space, after spontaneous gauge symmetry breaking, 
the relevant Lagrangian in the unitary gauge is given by
\bea
{\cal L}&=& {\cal L}_{F}\, +\, {\cal L}_V \, 
          +\, {\cal L}_H \, + \, h.c. \nonumber\\
{\cal L}_F&=& \sum_{i,j} \bar{f}_i\,\left( i\gamma^{\mu}\, D_{\mu}^{ij}\,-
M_i\,\delta^{ij}\right)\,f_j \nonumber\\
{\cal L}_V&=&-\frac{1}{4} F_{\mu\nu}(V) \, F^{\mu\nu}(V)\,+\,
\frac{M_V^2}{2}\,V_{\mu}\,V^{\mu}\nonumber\\
{\cal L}_H&=&\frac{1}{2}\left(\partial_{\mu} H \, \partial^{\mu} H -
M_H^2\,H^2\right)\,+\, \sum_i \lambda_i \left(\bar{f}_i\,f_i \, H\right)\,+\,
\frac{g}{2}M_V \,(V_{\mu}V^{\mu}\,H)
\label{Lag}
\eea
where
\bea
D_{\mu}^{ij}&=&\frac{1}{2} \dedouble_{\mu}\delta^{ij}-ig\,\left(g_V
+g_A\gamma_5\right) K^{ij} \, V_{\mu}\nonumber\\
F_{\mu\nu}(V)&=&\partial_{\mu} V_{\nu}-\partial_{\nu} V_{\mu}
\label{koba}
\eea
$\dedouble_{\mu}\,\equiv\,\deR_{\mu}-\deL_{\mu}$,
and $g_{V,A}$ represent axial and vectorial couplings
(e.g., in the $W$ case $g_{V}=-g_{A}=\frac{1}{2\sqrt{2}}$).  $K_{ij}$
is a unitary matrix which in this case generalizes the usual
CKM matrix.
Notice that we have restricted our Lagrangian to describe
only abelian gauge bosons, since we will consider processes 
involving at most two gauge bosons in each interaction.
The standard $W$ and $Z$ couplings to the Higgs boson and fermions
can be easily recovered by this lagrangian.

In order to obtain the expression for the energy momentum tensor 
$T_{\mu\nu}$ in eq.(\ref{gravity}), it is useful to rewrite 
${\cal L}$ in general space-time coordinates with the
metric $g^{\mu\nu}$. 
As usual, when there are fermion fields, this is simply achieved by 
the following standard procedure.
The Minkowski metric $\eta^{\mu\nu}$ is replaced
by the general metric $\eta^{\mu\nu} \to g^{\mu\nu}$ expressed in terms 
of the Vierbein fields $e_{a}^{~\mu}$ 
(i.e., $g^{\mu\nu}=\sum_a e_{a}^{~\mu} e_{a}^{~\nu}$, where
$a$ and $\mu, \nu$ are the Minkowski and world indices, respectively)
inside Eq.(\ref{Lag}), and the lagrangian ${\cal L}$ is multiplied
by $ \sqrt{-g}$,
where $g$ is the determinant of $g^{\mu\nu}$.
Then, the expression for $T_{\mu\nu}$
can be derived by expanding $e_{a}^{~\mu}$ around the flat metric 
$\delta_{a}^{~\mu}$
\beq
e_{a}^{~\mu}=\delta_{a}^{~\mu} + \frac{1}{\bar{M}_P} h_a^{~\mu}\, .
\eeq
At the first order in the $h^{\mu\nu}$ expansion, $T_{\mu\nu}$ is given by 
\bea
T_{\mu\nu}&=& \frac{i}{2}\bar{f}_i\left(\gamma_{\mu}D^{ij}_{\nu}
+\gamma_{\nu}D^{ij}_{\mu}\right)f_j\, -
\eta_{\mu\nu} \bar{f}_i \left(i\gamma^{\alpha}D^{ij}_{\alpha}
- M_i\delta^{ij}\,\right)f_j\,
\nonumber\\
&+& F_{\mu\alpha}\, F^{\alpha}_{~~\nu}\,+M_V^2 V_{\mu}V_{\nu}\,+
\frac{1}{2}\eta_{\mu\nu}\left(\frac{1}{2}F^{\alpha\beta}\,F_{\alpha\beta}
-\frac{1}{2} M_V^2 V_{\alpha} V^{\alpha}\right)
\nonumber\\
&+& \partial_{\mu} H\,
\partial_{\nu} H +g\,M_V\, V_{\mu}V_{\nu} H\,
-\frac{1}{2}\eta_{\mu\nu}\left(\partial_{\alpha} H
\partial^{\alpha} H -M_H^2\right)
\nonumber\\
&+&\eta_{\mu\nu} \left(\lambda_i\bar{f}_if_iH+
\frac{g}{2}\,M_V\,V_{\alpha}V^{\alpha}H\right)
\label{EMT}
\eea
where the sum over the $i,j$ fermion flavours is assumed.
Notice that, at first order in $h_{a}^{~\mu}$, 
there is no distinction between latin ($a$) and 
greek indices ($\mu$), being
all the contractions performed by $\eta_{\mu\nu}$, and 
$g^{\mu\nu}=\eta^{\mu\nu}+\frac{1}{\bar{M}_P}\left(
h^{\mu\nu}+h^{\nu\mu}\right)+ O(h^2)$.

By inserting eq.(\ref{EMT}) in eq.(\ref{gravity}), 
the corresponding Feynman rules for 
interaction vertices with both three--line and four--line vertices 
(and only one graviton emission) are easily obtained. 
We report their expressions
in Appendix A1.
All the four--line vertices contribute to the matrix elements involving 
on-shell
spin-2 fields except in the case of the fermion-Higgs couplings. In the
latter case, 
the $Hf\bar{f}G$ vertex is 
proportional to the trace of the energy momentum tensor
(see eq.(\ref{EMT})) and, therefore,
its spin--2 component vanishes.
\section{Heavy Gauge bosons and $Z$/$W$ decays}

\begin{figure}[tpb]
\dofig{3.1in}{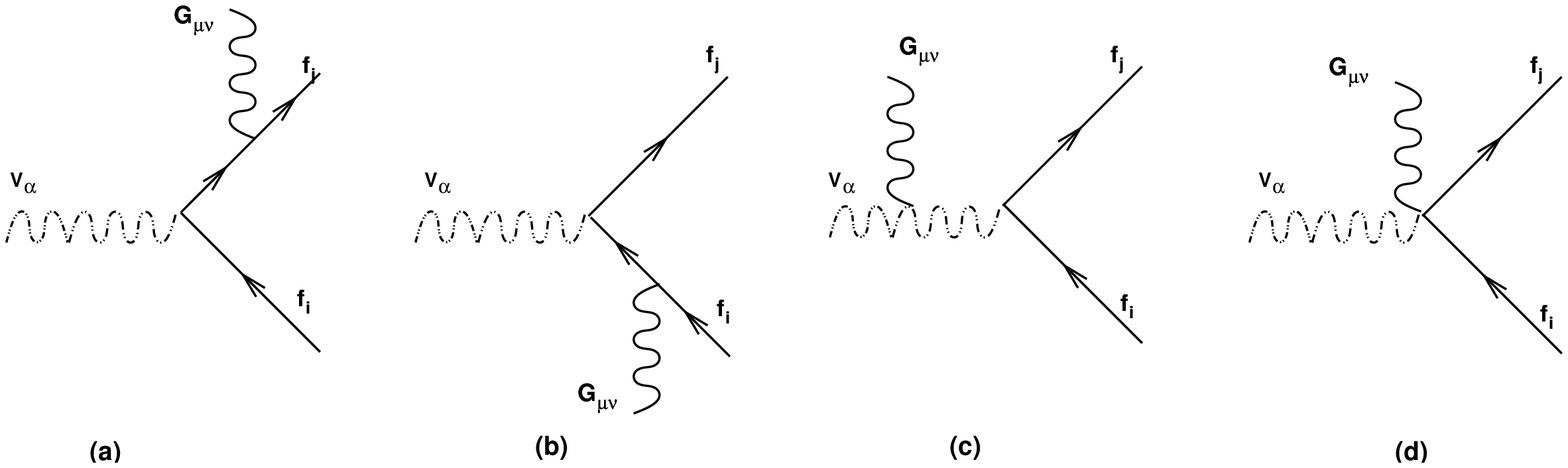 }
\caption{{\small Feynman diagrams (a-d) for the decays 
$V \to f \bar{f} + G$ or $f_i \to f_j V + G$. Analogous diagrams 
in the Higgs sector can be simply obtained by these diagrams, by 
replacing $V \to H$ for the decay $H \to f \bar{f} + G$, 
and also replacing $(f,\bar f)\to (V,V)$ for the decay $H\to V V + G$.
In the case $H \to f \bar{f} + G$ the (d) diagram vanishes, as explained 
in the text.
}}
\label{fig1} 
\end{figure}

We start our analysis by considering the following decay
\beq
V(p_V)\to \bar{f_i}(p_1)\,f_{i,j}(p_2)+G(p_G)
\eeq
where $V$ in this case can be both  a U(1) massive gauge boson and a
non-abelian SU(N) massive gauge boson,
$G$ is a massive spin-2 field, and $f_i$ is a generic fermion field.
We also set $K_{ij}=\delta_{ij}$ in eq.(\ref{koba}).
The four Feynman diagrams relevant to this process 
are shown in figs. \ref{fig1}a-\ref{fig1}d. 
We see
that the diagrams in figs. 1a-1c are obtained by attaching the spin-2 field 
in all possible 
ways to the external legs of the main diagram $V\to \bar{f}\, f$, while
fig. 1d is given by the contact term $VVHG$.

Since we are interested in studying unpolarized processes, we
recall here the formula for the sum over 
polarizations in the case of a massive  spin-2 field. This is given by
\cite{veltman}
\beq
\sum_{\sigma=1}^{5} \epsilon_{\mu\nu}(k,\sigma)\, \epsilon_{\alpha\beta}(k,\sigma)=
P_{\mu\nu\alpha\beta}(k)
\eeq
\bea
P_{\mu\nu\alpha\beta}(k)&=&\frac{1}{2}\left(\eta_{\mu\alpha}\eta_{\nu\beta}
+\eta_{\mu\beta}\eta_{\nu\alpha} -\eta_{\mu\nu}\eta_{\alpha\beta}\right)
\nonumber\\
&-&\frac{1}{2\, m_G^2}\left(\eta_{\mu\alpha}k_{\mu}k_{\beta}+
\eta_{\nu\beta}k_{\mu}k_{\alpha}+
\eta_{\mu\beta}k_{\nu}k_{\beta}+
\eta_{\nu\alpha}k_{\mu}k_{\beta}\right)
\nonumber\\
&+&\frac{1}{6}\left(\eta_{\mu\nu}+\frac{2}{m_G^2}k_{\mu}k_{\nu}\right)
\left(\eta_{\alpha\beta}+\frac{2}{m_G^2}k_{\alpha}k_{\beta}\right) ,
\label{pol}
\eea
where $\epsilon_{\mu\nu}(k,\sigma)$, $m_G$ and $k$ are the 
polarization tensor,  mass and momentum of the 
spin-2 field, respectively, 
and the index $\sigma$ runs over the polarization states.
Note that the projector $P_{\mu\nu\alpha\beta}$, which is 
symmetric and traceless in both $(\mu,\nu)$ and $(\alpha,\beta)$ indices,
satisfies the transversality conditions
$k^{\mu} P_{\mu\nu\alpha\beta} = k^{\alpha} P_{\mu\nu\alpha\beta}=0$.
Then, by using the Lagrangian in eqs.(\ref{gravity}) and 
(\ref{EMT}), 
the square modulus of the amplitude ${\cal M}$
summed over all final polarizations and averaged over the initial ones, 
is 
\beq
\frac{1}{3}\sum_{\rm pol} {|{\cal M}|^2}=
N_f
\frac{2\,g^2\,M_V^2}{3\,\bar{M}_P^2}
\left(
\left(|g_V|^2+|g_A|^2\right)\, F^{(+)}_V(t,u) + 
\left(|g_V|^2-|g_A|^2\right)\, F^{(-)}_V(t,u)
\right) 
\label{Vm2}
\eeq
where $g$ is the gauge coupling constant,
and $g_{V,A}$ are the vectorial and axial couplings defined in eq.(\ref{Lag}).
For fermion $f$,
$N_f$ represents the sum over quantum numbers, 
whose generators commute with the 
gauge group generator associated to the vector $V$,
like, for instance, the fermion color number 
in the case of realistic $Z$ or $W$ decays. 
We assume the two fermion masses degenerate (i.e., $M_{f_i}=M_{f_j}=M_{f}$). 
Then, we define the 
Mandelstam variables $t$ and $u$  as
\beq
t=\frac{1}{M_V^2}\,(p_1+p_G)^2-x_f,~~~
u=\frac{1}{M_V^2}\,(p_2+p_G)^2-x_f,~~~s=x_G-t-u
\eeq
where $x_f=\frac{M_f^2}{M_V^2}$, $x_G=\frac{m_G^2}{M_V^2}$,
and $M_V$, $m_G$, $M_f$ are the masses 
of the gauge boson, graviton, fermion, respectively.
The analytical expressions for the functions $F^{(\pm)}_V(t,u)$ 
(also depending on the variables $x_f$ and $x_G$) can be found in the
appendix A2. 

It is worth noticing that, despite the 
presence of $1/m_G^2$ terms in the sum over polarizations for the
massive spin-2 fields,
in the final expression for $F^{(\pm)}_V(t,u)$ 
(and analogously for the other decay functions in the appendix A2)
$m_G$
appears only with positive powers.  
The cancellation of $1/m_G^2$ terms in the total amplitude
is indeed ensured by the conservation (at the zeroth order in $h_{\mu\nu}$)
of the on-shell matrix elements of the 
energy momentum tensor in eq.(\ref{EMT}).
Therefore, the terms proportional to $k_{\mu}$ in eq.(\ref{pol}) 
do not contribute to the final amplitude.
As a severe check of our 
results\footnote{The functions appearing in 
appendix A2 were computed by FORM\cite{form}.}, 
we used the complete expression for
$P_{\mu\nu\alpha\beta}$ in eq.(\ref{pol}) and explicitly
verified this property.

Although the limit for $m_G\to 0 $ is smooth, our results for the
square amplitudes summed over polarizations is not supposed to coincide 
in this limit with the massless graviton contribution.
This is due to the well-known van Dam-Veltman discontinuity \cite{veltman}.
Our results only hold for $m_G\neq 0$. Indeed, the emission of a 
massless graviton should be calculated by using the 
proper massless projector  (see, e.g., \cite{veltman}),
that differs from the massive one.\footnote{
In particular,  $P_{\mu\nu\alpha\beta}(k)$ 
in eq.(\ref{pol}) for the massless case becomes \cite{veltman}
\beq
P_{\mu\nu\alpha\beta}^{\rm G}(k)=\frac{1}{2}\left(
\eta_{\mu\alpha}\eta_{\nu\beta}+\eta_{\mu\beta}\eta_{\nu\alpha}-
\eta_{\mu\nu}\eta_{\alpha\beta}
\right) + \dots
\label{pol_G}
\eeq
where the dots stand for any term containing at least one 
graviton momentum.
Notice that, also in this case, the terms proportional to the graviton momentum 
give zero when contracted with the on-shell matrix elements 
of $T_{\mu\nu}$, due to the conservation of the energy-momentum tensor.
Therefore, the  discontinuity in the limit  $m_G\to 0$ 
arises from terms  that do not 
contain the graviton momentum in the two  projectors.
In particular, the only difference in the relevant terms 
of eqs.(\ref{pol_G}) and (\ref{pol}) is given by the  
coefficients of 
$\eta_{\mu\nu}\eta_{\alpha\beta}$, that, when contracted with 
the on-shell matrix elements of the energy momentum tensor, 
give terms proportional to the trace $T_{\mu}^{\mu}$. 
Since we are analyzing massive
particles, the trace of $T_{\mu\nu}$ does not vanish, so in the limit 
$m_G\to 0$ we should expect a discontinuity.}
For the purpose of our analysis the effect of not
taking into account this discontinuity is
not relevant, since the square modulus of the amplitude for 
a single massless graviton contribution is suppressed by $1/\bar{M}_P^2$.

As said above, we are interested in analyzing 
inclusive 
processes, where one sums over all the kinematically allowed KK graviton states.
The mass splitting $\Delta m_G$ 
between different excitations is given by
\beq
\Delta m_G \sim \frac{1}{R} = 
M_D\left(\frac{M_D}{\bar{M}_P}\right)^{2/\delta}
\eeq
(e.g., in the case  $\delta< 4$, for $M_D\sim $1 TeV, 
$\Delta m_G$ is less than a few  KeV's).
This allows one to
approximate the KK modes as a continuous spectrum, with a number density
of modes $(d N)$ between $m_G$ and $m_G + d m_G$ given by \cite{GRW}
\beq
d N = S_{\delta-1} \frac{\bar{M}_P^2}{M_D^{2+\delta}} m_G^{\delta -1} d m_G .
\label{sum}
\eeq
Here, $S_{\delta-1}$ is the surface of a unit-radius sphere in 
$\delta$ dimensions.\footnote{From \cite{GRW}, we get
$S_{\delta-1}=2\pi^n/(n-1)!$ and  
$S_{\delta-1}=2\pi^n/\prod^{n-1}_{k=0} (k+\frac{1}{2})$ for
$\delta=2n$ and $\delta=2n+1$, with $n$ integer, respectively.}
Then, the integration over the number of KK states 
cancels the factor
$1/\bar{M}_P^2$  of the single graviton emission.

Finally, the result for the inclusive total width $\Gamma(V\to f\bar{f}+G_X)$, 
where $G_X$ indicates any KK graviton excitation up to the $M_V$ scale, 
is given by
\bea
\Gamma(V\to f\bar{f}\,+\,G_X)
&=&
N_f
\frac{M_V^3\, G_V\,S_{\delta-1}}{96\,\pi^3\,\sqrt{2}}
\left(\frac{M_V}{M_D}\right)^{2+\delta} \left(
\left(|g_V|^2+|g_A|^2\right)I^{(+)}_V(x_\Delta,x_f,\delta) \right. \nonumber\\
&+&
\left.
\left(|g_V|^2-|g_A|^2\right)I^{(-)}_V(x_\Delta,x_f,\delta)\right)
\label{Vwidth}
\eea
where 
\beq
I_V^{(\pm)}(x_{\Delta},x_f,\delta)=
\int_{x_{\Delta}}^{(1-2\sqrt{x_f})^2}\,
d\,x_G\,\,\left(x_G\right)^{\frac{\delta}{2}-1}\,
\int_{x_G+2 \sqrt{x_fx_G}}^{1-2\sqrt{x_f}}\,d\,t 
\int_{u_{-}}^{u_{+}}\,d\,u\,
F_V^{(\pm)}(t,u)
\label{int_V}
\eeq
and
\bea
u_{\pm}&=&\frac{\left(-2x_f+x_G-t\right)t+x_G+t
\pm \Delta}{2\left(t+x_f\right)}\nonumber \\
\Delta&=&\sqrt{\left(\left(x_G-t\right)^2-4x_fx_G\right)
\left(1+t^2-2\left(2x_f+t\right)\right)},~~~
x_{\Delta}=
\frac{\Delta_{\rm exp}^2}{M_V^2}
\label{Vdecay}
\eea
We defined $G_V=g^2/(4\,\sqrt{2}\, M_V^2)$ extending the definition of the
standard Fermi constant $G_F$. $\Delta_{\rm exp}$ is the experimental
resolution on the invariant mass of missing energy.
Notice that the function $I_V^{(-)}(x_{\Delta},x_f,\delta)$,
that is proportional to  the fermion masses, vanishes
in the $x_f \to 0$ limit. The integral over the 
phase space is also regular in the 
$x_{\Delta} \to 0$ limit. This property is connected to
the absence of graviton mass 
singularities in the square amplitude.

In table 1, we report some numerical values of the 
integrals $I_V^{(+)}(x_\Delta,x_f,\delta)$ in eq.(\ref{int_V}), 
evaluated at $x_\Delta=0$ and
$x_f=0$, for the representative cases $\delta=2,3,\dots,6$.
\begin{table}
\begin{center}
\begin{tabular}{|r||c|c|c|c|}
\hline 
${\rm \delta}$
& $I_V^{(+)}$ & $ I_f$ & $ I_H^f$ & $I_H^V$ \\ \hline \hline 
$2$ & $1.20\times 10^{-1}$   & $5.53\times 10^{-2} $ & $3.25\times 10^{-1} $ & $7.78\times 10^{-2}$ \\ \hline
$3$ & $2.12\times 10^{-2} $  & $8.72 \times 10^{-3}  $ & $4.95\times 10^{-2} $ & $1.17\times 10^{-2}$ \\ \hline
$4$ & $5.85 \times 10^{-3}$  & $2.10\times 10^{-3} $ & $1.14\times 10^{-2} $ & $2.67\times 10^{-3}$ \\ \hline
$5$ & $2.11 \times 10^{-3}$ & $6.57 \times 10^{-4} $ & $3.35\times 10^{-3} $ & $7.86\times 10^{-4}$ \\ \hline
$6$ & $9.13 \times 10^{-4}$ & $2.45 \times 10^{-4}$ & $1.18\times 10^{-3}  $ & $2.76\times 10^{-4}$ \\ \hline
\end{tabular}
\caption[]{Numerical values for 
$I^{(+)}_V(x_\Delta,x_f,\delta),~I_f(x_\Delta,x_V,\delta),
~I_H^f(x_\Delta,x_f,\delta)$, and 
$I_H^V(x_\Delta,x_V,\delta)$ 
evaluated at $x_\Delta=x_V=x_f=0$, and for  
$\delta=2,3,4,5,6$. }
\label{INT}
\end{center}
\end{table}
In fig. \ref{fig2}, we plot the integrals
$I_V^{(+)}(x_\Delta,x_f,\delta)$ (evaluated at $x_\Delta=0$
and divided by their values at $x_f=0$ [see table \ref{INT}])
versus $x_f$, and for $\delta=2,3,\dots,6$.
In fig. \ref{fig3}, we plot the differential 
widths versus the ratio  $r_G=m_G/M_V$, 
and for $\delta=2,3,\dots,6$. In particular, we plot the
distribution  $R$, normalized  as 
\beq
R  =
\frac{1}{
\frac{d \Gamma}{d r_G}|_{{\rm max}}}
~\frac{d \Gamma}{d r_G} 
\label{width}
\eeq
where $\frac{d \Gamma}{d r_G}|_{{\rm max}}$ stands for the  maximum 
of $\frac{d\Gamma}{d r_G}$ versus $r_G$.
The shape of this distribution provides information on the typical 
fraction of missing energy, due to
the KK gravitons emission,
expected in the decay.
We analyze two representative cases: $M_f = 0$
and $M_f = 0.2~M_V$.
From these results, we see that the position of the maximum ($r^{max}$) of the
distributions $R$ is quite sensitive to 
the number of extra dimensions, going from $r^{\rm max}\simeq 0.1$, 
for $\delta=2$, up to $r^{\rm max}\simeq 0.5$, for $\delta=6$.
When the mass for the final fermion is taken into account,
 these curves shift toward lower  $r_G$ values,
due to the phase-space reduction of the allowed $x_G$ 
range.

We recall that our perturbative treatment is bound to be valid for decaying
particles not heavier than $M_D$.
Indeed, being gravity  directly 
coupled to the energy momentum tensor, the validity of the perturbative
expansion strongly depends on the energy scale of
the process with respect to the 
Planck mass $M_D$. Since in the relevant scenarios $M_D$ could be 
close to the TeV scale, this question is not just academic.
In  \cite{GRW}, 
when considering direct graviton production at colliders, 
upper bounds on the
center of mass energy as a function of $M_D$ and number of extra dimensions
have been obtained 
by requiring unitarity of the tree-level cross sections for 
single graviton production.
In our case, if new particles exist with masses either larger 
than or close to 
the fundamental Planck scale $M_D$, then
non-perturbative gravitational 
phenomena should sizeably affect
their decay width. For instance,
mass upper bounds that somewhat limit 
the perturbative regime can be obtained by  requiring that 
the rate for one graviton emission from the main particle decay 
does not exceed the rate of its main decay. In particular, one can impose
\beq 
\frac{\Gamma(V\to f\bar{f}+G_X)}{\Gamma(V\to f\bar{f})} < 1
\label{Vbounds}
\eeq
for any fermion $f$ and for $\Delta_{\rm exp}=0$.
In the approximation of massless fermions the width of 
$V\to f\bar{f}$ is given by
\beq
\Gamma(V\to f\bar{f})= N_f \frac{2\,M_V^3\, G_V}{3\,\pi\,\sqrt{2}}
\left(|g_V|^2+|g_A|^2\right)
\label{Vtreedecay}
\eeq
then from Eq.(\ref{Vbounds}), one obtains
\beq
M_V < M_D \left(\frac{64\,\pi^2}{I^{(+)}_V(x_\Delta,0,\delta)\,S_{\delta-1}}
\right)^{\frac{1}{2+\delta}}\,.
\label{boundsMV}
\eeq
From the  values of the integrals given in table 1,
for $x_\Delta=0$ we find
\beq
M_V < M_D \times \left( 5.4,~4.7,~4.2,~3.8,~3.5 \right),
\label{bounds_unV}
\eeq
for $\delta=2,3,4,5,6$, respectively.
For massive fermions,
the value of $I^{(+)}_V$  would be smaller (due to a smaller
available phase space), giving less stringent upper bounds on $M_V$.

As a phenomenological application of this study, 
we analyze now the constraints on $M_D$
which come from negative searches of extra missing energy events in the
$Z$ decays. 
In particular, we will consider the process
\beq
Z \to \,\bar{f}\, f \,+\, G_X
\eeq
where $f$ may indicate leptons and quarks.
The massless limit for 
the final fermion states is quite accurate in this case.
From eq.(\ref{Vtreedecay}), the decay width of $Z\to f_i \bar{f}_i$
for massless fermions is simply  
\beq
\Gamma(Z\to f\bar{f})=
N_f \frac{2\,M_Z^3\, G_F}{3\,\pi\,\sqrt{2}}
\left(|g_V^f|^2+|g_A^f|^2\right)
\eeq
where $N_f=1$ and 3 for leptons and quarks respectively, and
$g^f_A=T^f_3/2$, 
$g^f_V=\left(T_3^f-2Q^f\sin{\theta_W}^2
\right)/2$, being $T_3^f$ and $Q^f$
the eigenvalues of the third component of isotopic spin ($T_3^e=-\frac{1}{2}$)
and electric charge, respectively.
Here, the vectorial and axial couplings  have been properly normalized
after the introduction of  the true Fermi constant $G_F$.
Then, the branching ratio (BR) for the inclusive decay 
$Z\to \sum_f f\bar{f}+G_X$, where 
$f$ stands for any quark or lepton in the final state, is given by
\beq
{\rm BR}(Z\to \sum_f f\bar{f}+ G_X)=\frac{S_{\delta-1}}{64\,\pi^2}
\left(\frac{M_Z}{M_D}\right)^{2+\delta} \, I_V^{(+)}(x_{\Delta},0,\delta)
\label{BRZ}
\eeq
In the case  $\delta=2$, and for $x_{\Delta}=x_f=0$, we obtain
\beq
{\rm BR}(Z\to \sum_f f\bar{f}+ G_X)=
\frac{8.2\times 10^{-8}}{M^4_D({\rm TeV})}
\label{BR_Z_2}
\eeq
The result in eq.(\ref{BR_Z_2})  agrees 
with the corresponding one in \cite{Zpole}, 
after identifying  
$M^4_{\star}=2\,M_D^4$ for $\delta=2$, being 
the definition of $M_{\star}$ in \cite{Zpole}
different from $M_D$ in eq.(\ref{newton})
($G_N^{-1}= 4 \pi R^{\delta} M_{\star}^{2+\delta}$).
Note that in \cite{Zpole} some graviscalar contribution is included.

These results can be extended to the decay 
$W^{\pm}\to f f^{\prime} +G$, in the massless
limit for final fermions.
In the latter case, the width and inclusive BR 
can be simply obtained from the $Z$ case, 
by replacing $M_Z \to M_W$ in eq.(\ref{BRZ}).
In the case $\delta=2$, and for $x_{\Delta}=0$, we obtain
\beq
{\rm BR}(W^{\pm}\to \sum_{f,f^{\prime}} f^{\prime}\bar{f}+ G_X)=
\frac{5.0\times 10^{-8}}{M^4_D({\rm TeV})}
\label{BR_W_2}
\eeq
where $f\neq f^{\prime}$ run over the leptons and quarks  
with $T_3^f=1/2$ and $T_3^{f^{\prime}}=-1/2$.

The result in eq.(\ref{BR_Z_2}) can be applied
 to the LEP1 data on $Z\to f\bar{f}+ E_{miss}$, corresponding to
 about $2\times 10^7$ $Z$ decays. 
 The SM background, given by the four-fermion decay
$Z\to f\bar{f}\nu\bar{\nu}$, is very small. 
A few events that are quite in agreement with the SM prediction 
were observed \cite{kobel}.
Assuming, one can push the limit on unexpected signals down to
 BR($Z\to \sum_f f\bar{f}+ G_X)<10^{-7}$, one then gets,
for $\delta=2$, 
\beq
M_D > 951\,{\rm GeV} 
\label{boundsZ}
\eeq
This limit is not far from what is obtained from 
the negative searches at LEP2 in the channel
$e^+e^-\to \gamma \,+\, E_{miss}$, (where $E_{miss}$ is the 
missing energy due to gravitons emission) and from virtual gravitons
effects \cite{land}. 

\section{Heavy Fermions and Top decays}
We consider now  a heavy fermion $f_i$ decaying into a lighter fermion 
$f_j$ plus a massive vector boson and a graviton
\beq
f_i(p_i)\to f_j(p_j)\, V(p_V) + G(p_G)
\label{topd}
\eeq
This class of processes  includes the
 decay of the top quark $t\to W b +G$ in the SM, that we will
discuss later on.
By summing over all final polarizations and averaging over initial ones, 
the square modulus of the amplitude for the process (\ref{topd})
is given, in the massless limit for $f_j$, by
\beq
\frac{1}{2}\sum_{\rm pol} {|{\cal M}|^2}=
\frac{g^2\,M_f^4\,|K_{ij}|^2}{\bar{M}_P^2\,M_V^2}
\left(|g_V|^2+|g_A|^2\right)\, F_f(t,u)
\label{Fm2}
\eeq
where the expression for the function $F_{f}(t,u)$ 
can be found in the appendix A2. Here, the Mandelstam variables 
are defined as
\beq
t=\frac{1}{M_{f_i}^2}\,(p_i-p_G)^2-1\,,~~~
u=\frac{1}{M_{f_i}^2}\,(p_i-p_V)^2\,,~~~
s=x_G-t-u\,,
\eeq
where $p_i^2=M_{f_i}^2,~~p_V^2=M_V^2$, 
$x_V\equiv \frac{M_V^2}{M_{f_i}^2}$ and $x_G\equiv \frac{m_G^2}{M_{f_i}^2}$.
By the same procedure explained in the previous section,
we got the total width $\Gamma(f_i\to f_j\,V\,+\,G_X)$ for the inclusive 
KK graviton production. In the massless limit for $f_j$, this is given 
 by
\beq
\Gamma(f_i\to f_j\, V\, +\,G_X)=
\frac{M_f^3\, G_V |K_{ij}|^2\,S_{\delta-1}}{64\,\pi^3\, \sqrt{2}}
\left(|g_V|^2+|g_A|^2\right)
\left(\frac{M_f}{M_D}\right)^{2+\delta}\,I_f(x_{\Delta},x_f,\delta)
\eeq
where $G_V=g^2/(4\,\sqrt{2}\, M_V^2)$, and
\bea
&&I_f(x_{\Delta},x_V,\delta)
= 
\int_{x_\Delta}^{\left(1-\sqrt{x_V}\right)^2} \,
d\,x_G\,\,\left(x_G\right)^{\delta/2-1}
\,\int_{x_G}^{\left(1-\sqrt{x_V}\right)^2}
\,d\,u \int_{t_{-}}^{t_{+}}\,d\,t\,F_f(t,u)
\nonumber \\
t_{\pm}&=& \frac{u-x_G}{2\,u}\left(1-u-x_V\right)\left(1\pm \Delta\right)
-1+x_V\,
~~~~\Delta=\sqrt{1-\frac{4 u x_V}{(1-u-x_V)^2}}\,
\eea
with $x_\Delta=\frac{\Delta_{\rm exp}^2}{M_{f_i}^2}$.

\begin{figure}[tpb]
\makefigs4{3.1in}{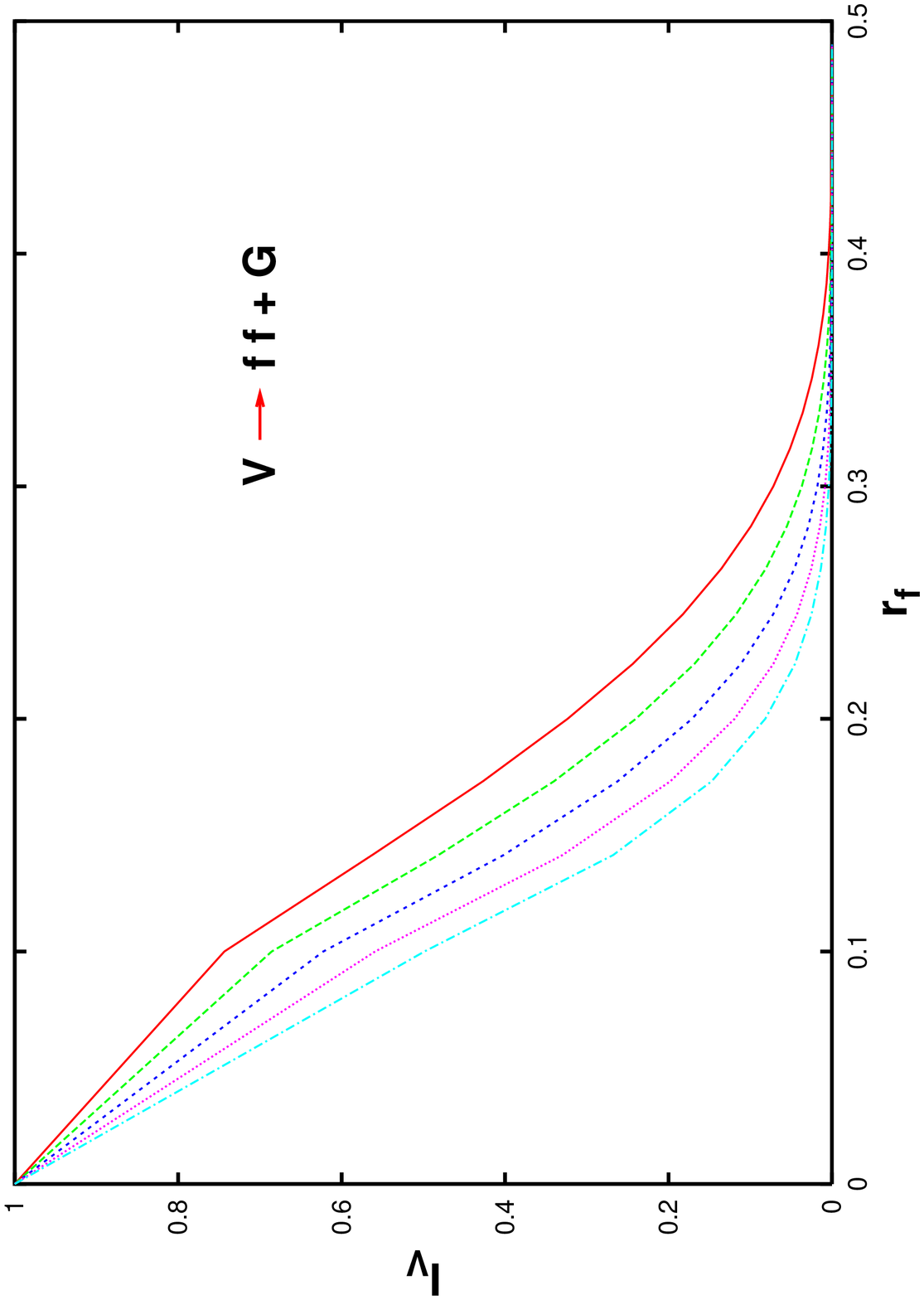 }
{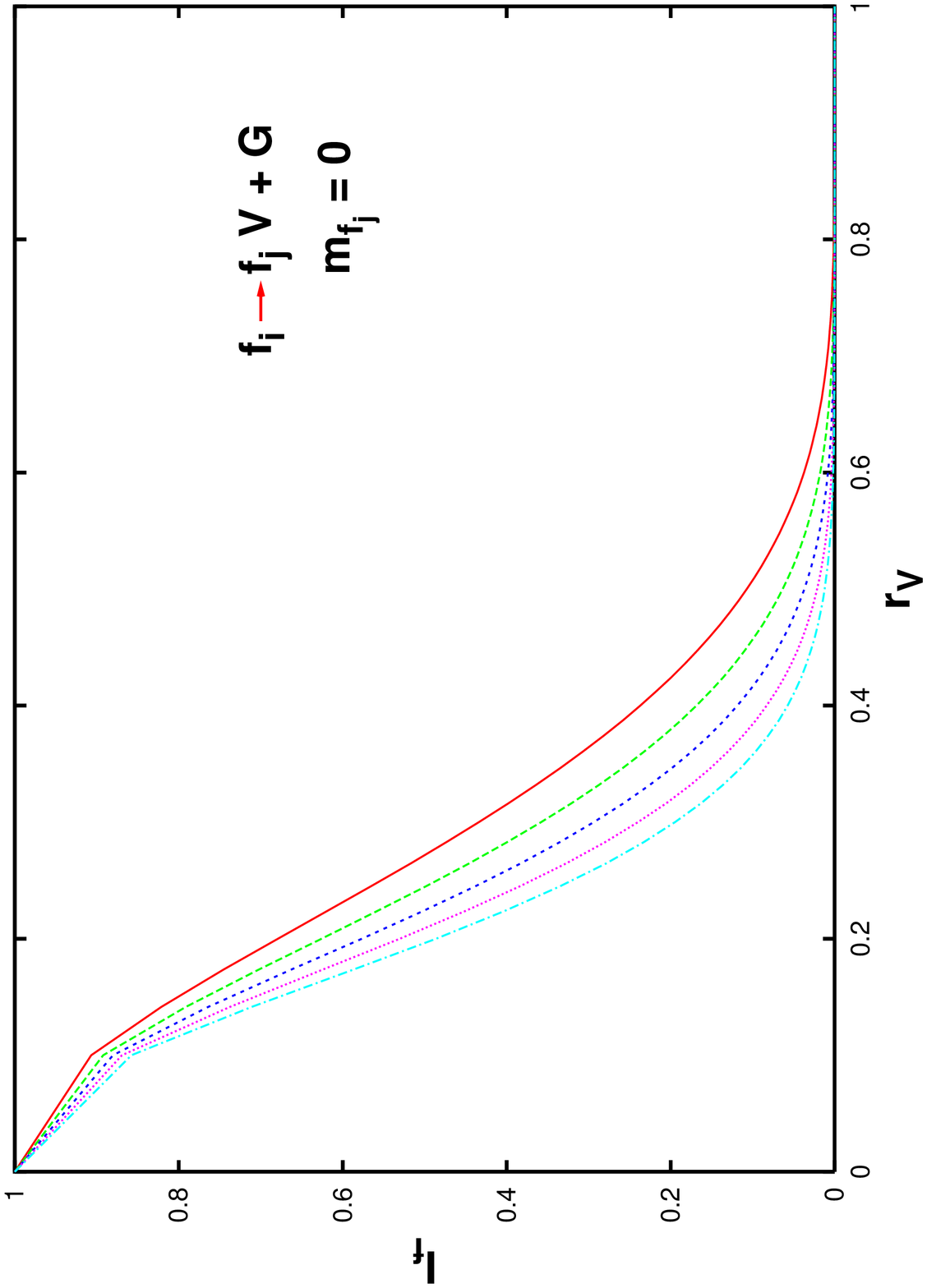 }{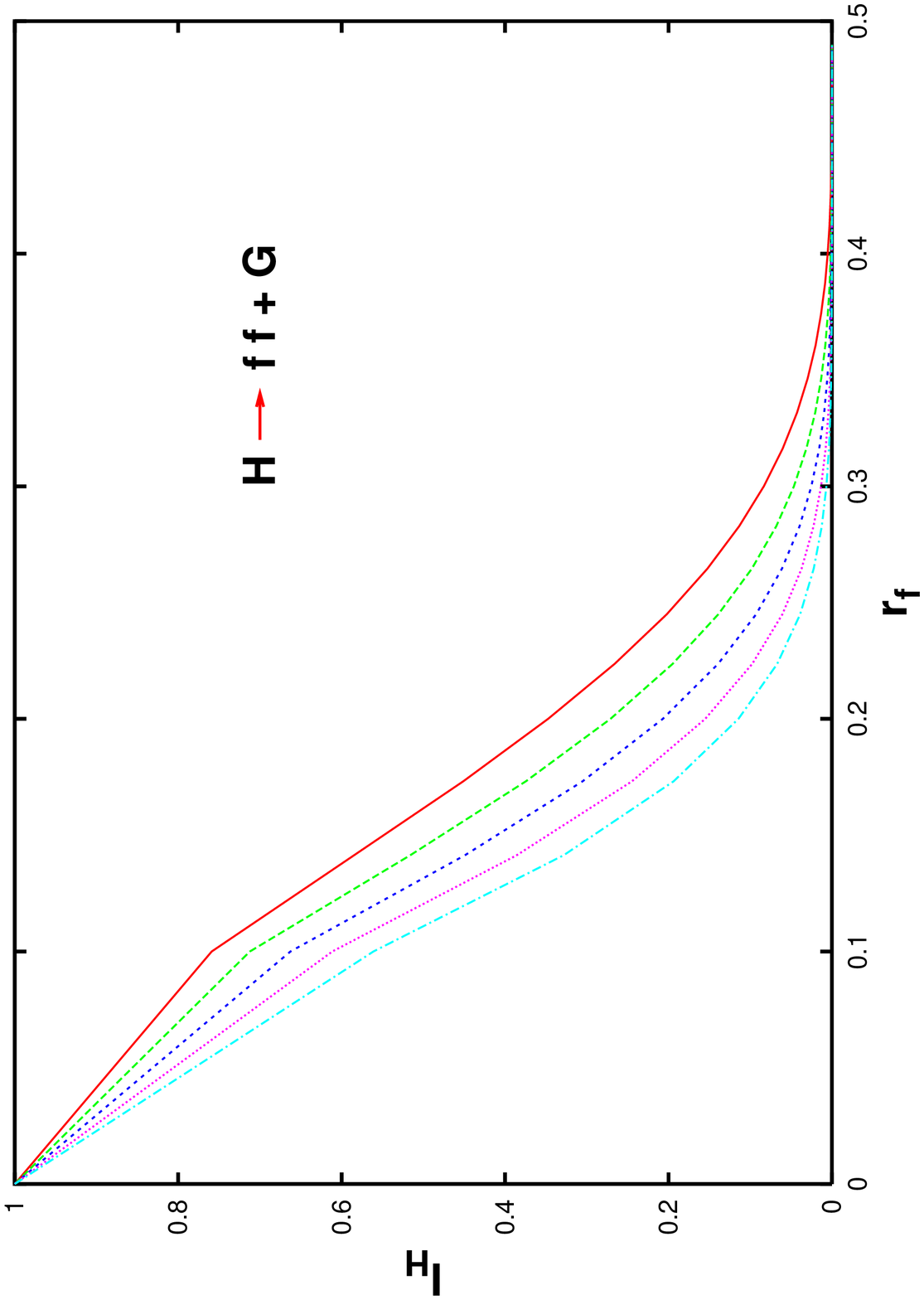 }{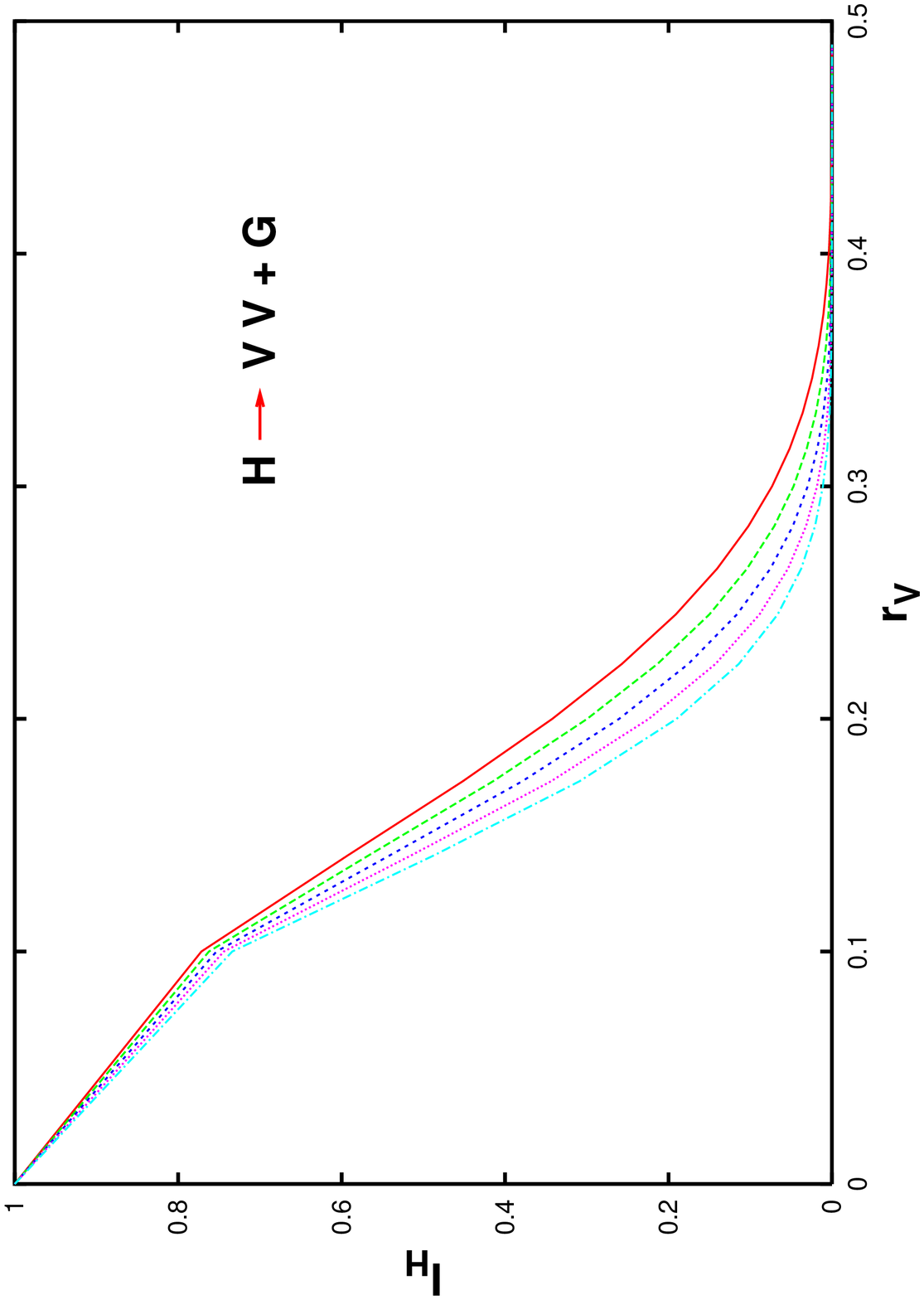 }
\caption{{\small Values of the integrals 
$I_V$ (top-left) , $I_{\rm F}$ (top-right), 
$I_H^f$ (bottom-left), and $I_H^V$ (bottom-right)
versus $r_V$, $r_{f_j}$, $r_{f}$, and $r_V$ respectively,
evaluated at $x_\Delta=0$, and divided by their corresponding 
values at $x_i=0$. A superscript $I^{(+)}$ is understood for $I_{V}$.
In each plot, lower curves 
correspond to higher  extra dimensions, for  $\delta=2,3,4,5,6$, respectively.
Here, $r_i=\frac{M_i}{M}$, $x_i=r_i^2$,
where $M$ is the mass of the decaying particle.
}}
\label{fig2} 
\end{figure}

\begin{figure}[tpb]
\makefigs4{3.1in}{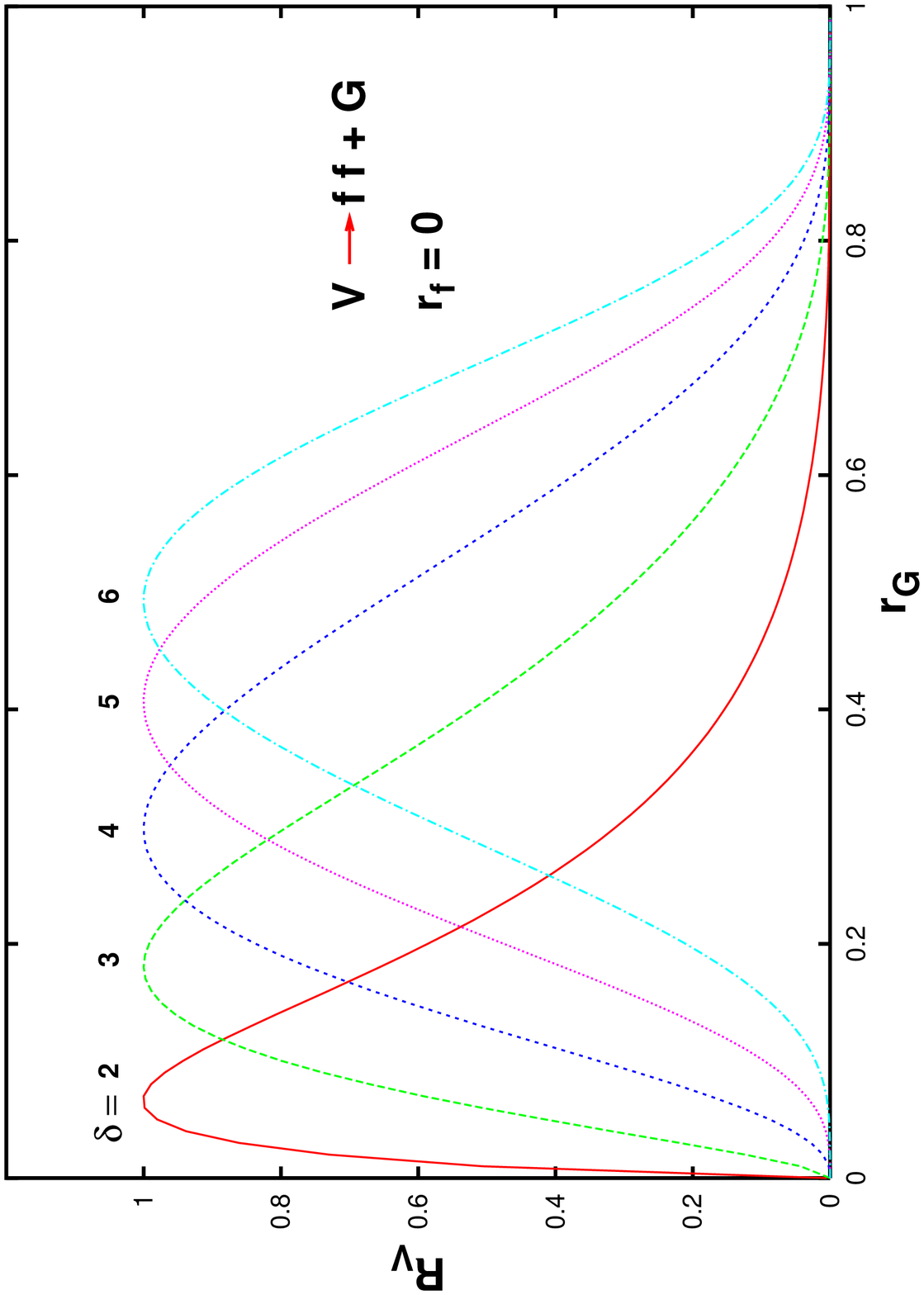 }{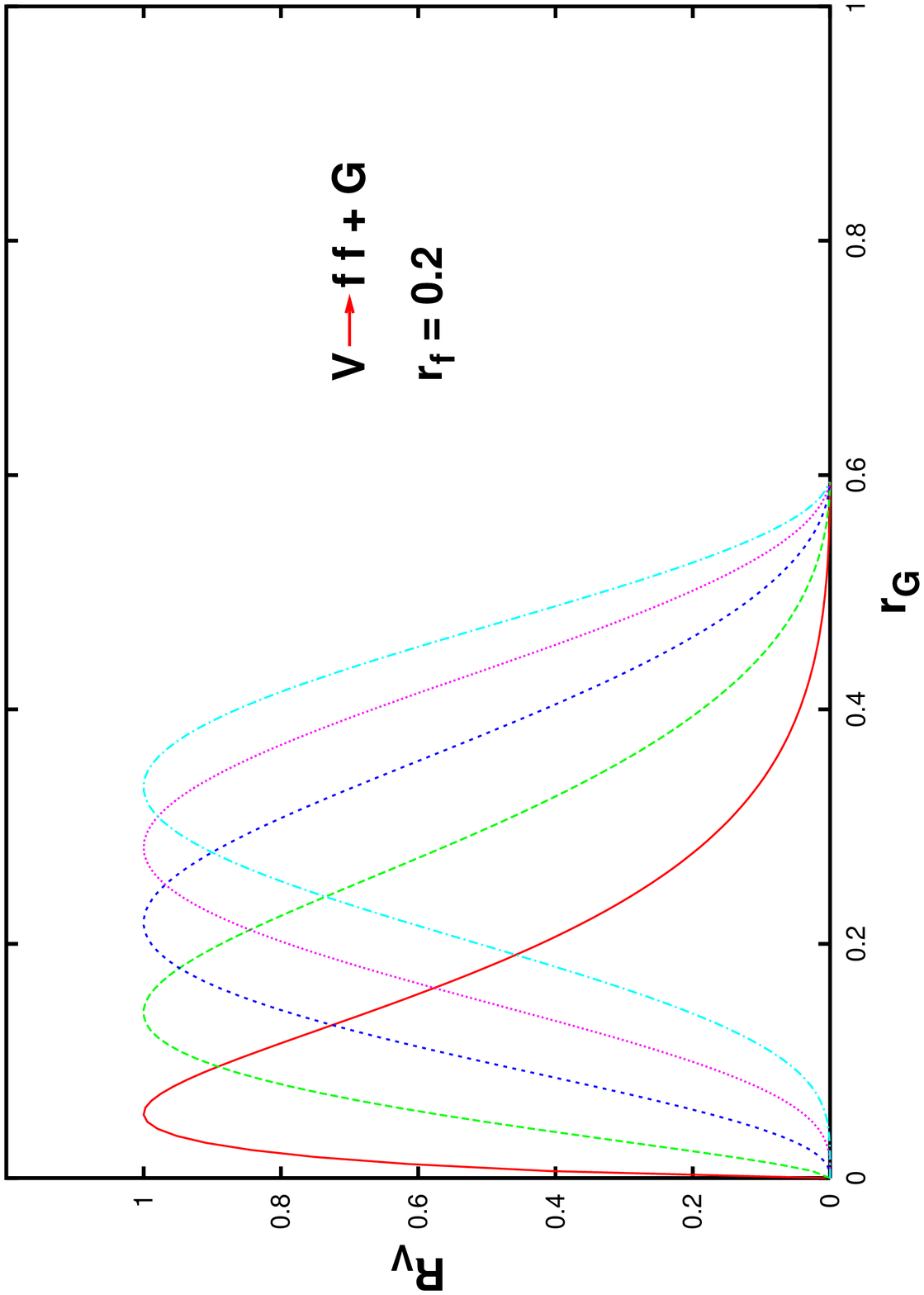 }
{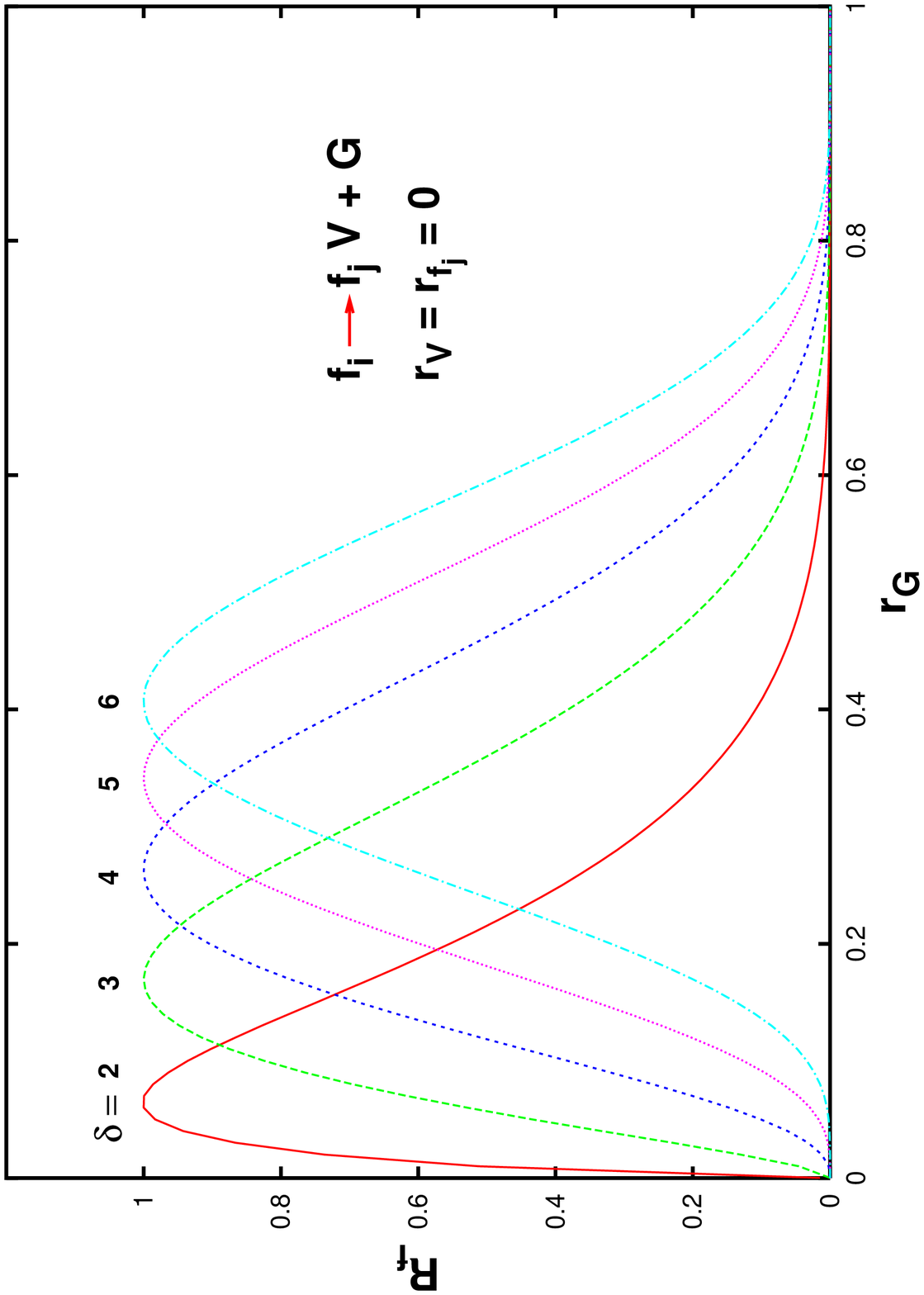 }{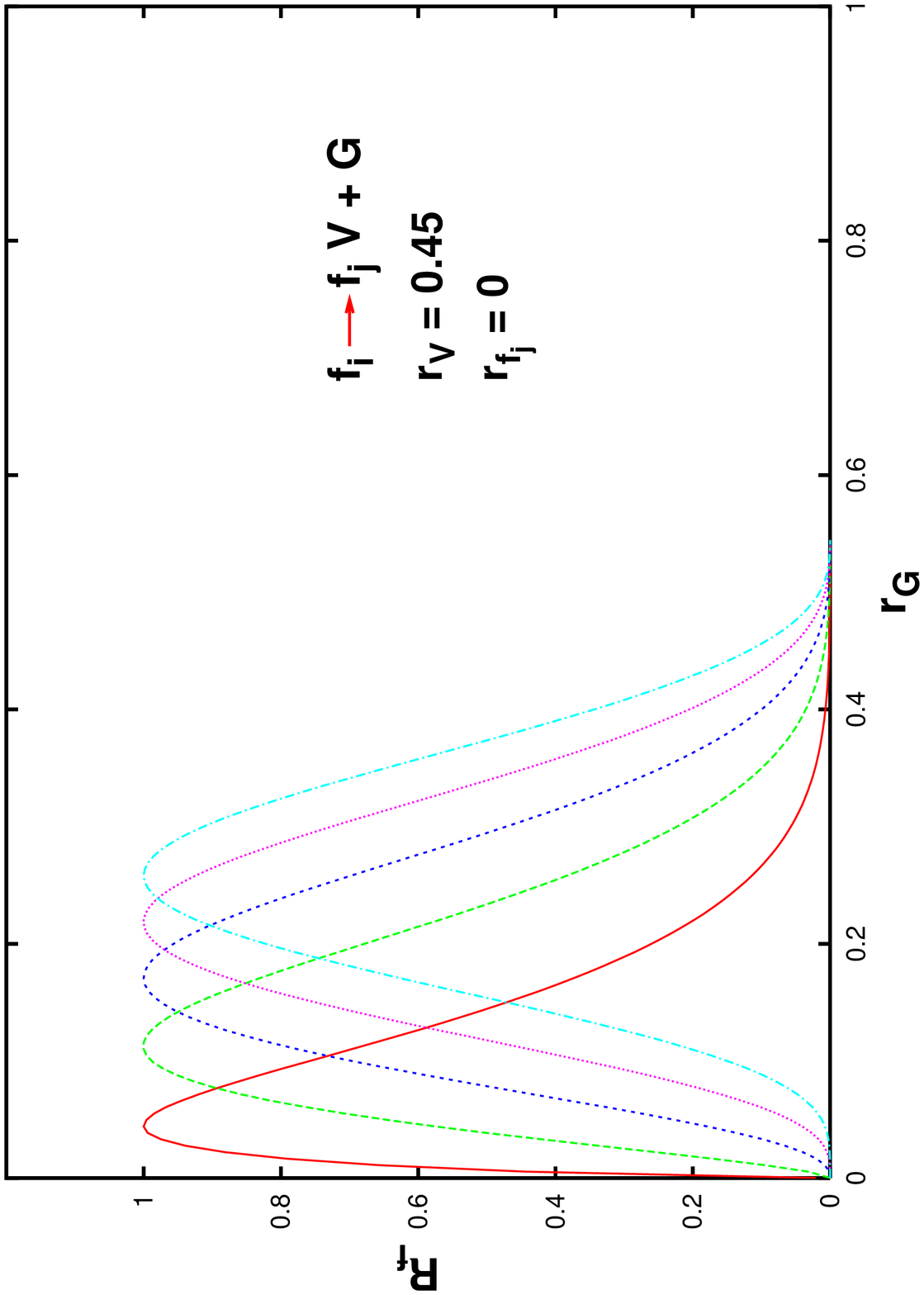 }
\caption{{\small Widths distributions 
$R=\frac{1}{
\frac{d \Gamma}{d r_G}|_{{\rm max}}}
~\frac{d \Gamma}{d r_G} $, 
versus $r_G$, for $x_\Delta=0$, and for  
$\delta=2,3,4,5,6$. Here, $r_G=\frac{m_G}{M}$,
$r_i=\frac{M_i}{M}$ with $i=f,V$ and
$M$ is the mass of the decaying particle. Plots relative to the
$V\to \bar{f} f +G$ and $f_i\to f_j V +G$ decays are shown.
}}
\label{fig3} 
\end{figure}

In table 1, we present 
the numerical results for $I_f(x_{\Delta},x_V,\delta)$, 
at $x_\Delta=x_V=0$,
and in fig. \ref{fig2} we plot the function 
$I_f(0,x_V,\delta)/I_f(0,0,\delta)$  versus $x_V$, 
 for representative values  $\delta=2,3,\dots,6$.
In fig. \ref{fig3}, we plot the differential 
widths defined in eq.(\ref{width})  versus  
$r_G\equiv \sqrt{x_G}$, 
and for $\delta=2,3,\dots,6$. We consider both 
the massless vector boson case and the massive case with 
$r_V =0.45$. This value of $r_V$ is relevant in the  top quark 
decay, where  $r_W=M_W/M_t\simeq 0.45$ ($M_t$ is the top quark mass).
By comparing the distributions of the vector boson and fermion decays in
fig. \ref{fig3}, we see that in general the positions of the 
maximum for the fermion distributions 
are closer to zero than in the vector boson case.

By requiring unitarity for the perturbative expansion,
\beq 
\frac{\Gamma(f_i\to f_j\,V+G)}{\Gamma(f_i\to f_j\,V)} < 1
\label{Fbounds}
\eeq
where the  total width $\Gamma(f_i\to f_j\,V)$  
(in the $f_j$ massless limit) is 
\beq
\Gamma(f_i\to f_j\, V\,)=\frac{M_{f_i}^3\,G_V |K_{ij}|^2}{2\,\pi\,\sqrt{2}}
\left(|g_V|^2+|g_A|^2\right)
\rho(x_V)
\eeq
and $\rho(x)=1-3x^2+ 2x^3$, we obtain
\beq
M_{f_i} < M_D \left(\frac{32\,\rho(x_V)\,\pi^2}
{I_f(x_{\Delta},x_V,\delta)\,S_{\delta-1}}
\right)^{\frac{1}{2+\delta}}\,.
\label{boundsMf}
\eeq
From the $I_f$ values  
 in table 1, we get for $x_V=x_{\Delta}=0$ 
the following limits from unitarity
\beq
M_f < M_D \times \left(5.5,~4.9,~4.4,~4.0,~3.8\right),
\label{bounds_unF}
\eeq
where the numbers inside parenthesis correspond to
$\delta=2,3,4,5,6$, respectively.

Now we apply these results to the specific case of the top quark decays, where 
$V=W^{\pm}$ and $g_V=-g_A=\frac{1}{2\sqrt{2}}$.
The top total width  
(at tree level, and neglecting CKM
nondiagonal decays) is, in the $b$ massless limit
\beq
\Gamma(t\to W\, b)=\frac{M_t^3\,G_F |V_{tb}|^2}{8\,\pi\,\sqrt{2}} \rho(x_W)
\eeq
where  $\rho(x_W)=0.887$, and $V_{ij}$ is the standard CKM matrix.
Then the total inclusive $BR$ for any $\delta$ is given by
\beq
{\rm BR}(t\to W\, b\,+\,G_X)=\frac{S_{\delta-1}}{32\, \pi^2}
\left(\frac{M_t}{M_D}\right)^{2+\delta} 
\frac{I_{f}(x_{\Delta},x_W,\delta)}{\rho(x_W)}
\eeq
In the case $\delta=2$, and $x_{\Delta}=0$, we obtain
\beq
{\rm BR}(t\to W b+ G_X)=
\frac{1.8\times 10^{-7} }{M^4_D({\rm TeV})}
\label{BR_Top_2}
\eeq
being $I_{f}(0,x_W,2)=8.38\times 10^{-3}$ (for $M_t=175$GeV).

It can be interesting to compare this value, with the rates expected
for other rare top quark decays both inside and beyond the standard
model, also considering the potential of future accelerators in this field
\cite{Beneke:2000hk}.
In  case of negative searches for this signal, one will
 impose an experimental upper bound  on the 
BR of this decay: 
${\rm BR}(t\to W\, b\,+\,G_X) < \Delta_{exp}^{\rm top} $, where 
$\Delta_{exp}^{\rm top}$ is related to the experimental sensitivity 
on the top branching ratio. Then, one finds, for $\delta=2$,
\beq
M_D > \left(\Delta_{exp}^{\rm top}\right)^{-\frac{1}{4}} \, 0.22\, M_Z 
\eeq
where we expressed the mass scale $M_t$ on  the right hand side
through $M_Z$. Then,
we can compare this result with the corresponding 
one for the Z decay, obtained from eq.(\ref{BRZ}) for $\delta=2$, 
\beq
M_D > \left(\Delta_{exp}^Z\right)^{-\frac{1}{4}} \, 0.19\, M_Z 
\eeq
where ${\rm BR}(Z\to  \bar f f\,+\,G_X) < \Delta_{exp}^Z$ is assumed.
We see that, assuming (at present, quite unrealistically) 
a comparable  sensitivity on the  two BR's, the lower bounds on $M_D$
obtained from the gravitational $Z$ and top decays turn out 
to be comparable, too.

\section{Higgs boson decays}
In this section, we  analyze
the gravitational emission in the Higgs boson decays 
into either two massive gauge
bosons  or  two fermions, 
\bea
H(p_H)&\to& V(p_1)\,\, V(p_2)\,+\,G(p_G)\nonumber\\
H(p_H)&\to& \bar{f}(p_1)\, f(p_2)+G(p_G) .
\eea
Using the interaction vertices given in section 2, we obtain 
the following expressions
for the
square modulus of the amplitudes summed over polarizations
\bea
\sum_{\rm pol} {|{\cal M}_{H\to V\, V\, +\,G}|^2}&=&
\frac{g^2\,M_H^4}{6\,\bar{M}_P^2\,M_V^2}\, F_H^V(t,u)\nonumber\\
\sum_{\rm pol} {|{\cal M}_{H\to \bar{f}\, f\, +\,G}|^2}&=&
N_f \frac{\lambda_f^2\,M_H^2}{3\,\bar{M}_P^2}\, F_H^f(t,u)
\label{Hm2}
\eea
where  the functions $F_{\rm H}^{V,f}(t,u)$ 
can be found in appendix A2.
The definition of Mandelstam variables $t,u$, and $s$ is given by
\beq
t=\frac{1}{M_H^2}\,(p_1+p_G)^2\, -\, x_{f,V},~~~
u=\frac{1}{M_H^2}\,(p_2+p_G)^2\, -\, x_{f,V},~~~
s=x_G-t-u
\eeq
with $x_i=\frac{M_i^2}{M_H^2}$, $i=V,F$.
Then, the inclusive total widths are given by
\bea
\Gamma(H\to V\, V\, +\,G_X)&=& \kappa
\frac{M_H^3\, G_V\, S_{\delta-1}}{384\,\pi^3\, \sqrt{2}}
\left(\frac{M_H}{M_D}\right)^{2+\delta} I_{\rm H}^V(x_{\Delta},x_V,\delta)
\label{hvv}\\
\Gamma(H\to \bar{f}\, f +G_X)&=& N_f
\frac{M_H\, \lambda_f^2\,S_{\delta-1} }{1536\,\pi^3}
\left(\frac{M_H}{M_D}\right)^{2+\delta} I^f_{\rm H}(x_{\Delta},x_f,\delta)
\eea
where
\beq
I_{\rm H}^{V,f}(x_{\Delta},x_{(f,V)},\delta)=
\int
d\,x_G\,\,\left(x_G\right)^{\delta/2-1}
\,\int\,d\,t \int \,d\,u\,F^{V,f}_{\rm H}(t,u).
\eeq
Here, the integration limits are the same as in
$V\to \bar{f} f +G$ (see eq.(\ref{int_V})), with 
$x_f\to x_V$ in the case of $I_H^V$.
The coefficient $\kappa$ in eq.(\ref{hvv}) is equal to 1, unless
the two final vector bosons are identical particles.
In the latter case, $\kappa=\frac{1}{2}$.

The tree level decay widths of $H\to \bar{f} f$ and $H\to V V $
are given by 
\beq
\Gamma(H\to \bar{f}\, f)= N_f
\frac{M_H\, \lambda_f^2}{8\,\pi}\rho_f(x_f),~~~~
\Gamma(H\to V\, V)= \kappa
\frac{M_H^3\, G_V}{8\,\pi\, \sqrt{2}}\rho_V(x_V)
\label{hvvdue}
\eeq
where $\rho_f=\left(1-4x\right)^{\frac{3}{2}}$ and 
$\rho_V(x)=\sqrt{1-4x}~(1-4x+12x^2)$.

The unitarity conditions here require 
\beq
M_{H} < M_D \left(\frac{192\,\rho_f(x_f)\,\pi^2}
{I_H^f(x_{\Delta},x_f,\delta)\,S_{\delta-1}}
\right)^{\frac{1}{2+\delta}}\, ,~~~~~
M_{H} < M_D \left(\frac{48\,\rho_V(x_V)\,\pi^2}
{I_H^V(x_{\Delta},x_V,\delta)\,S_{\delta-1}}
\right)^{\frac{1}{2+\delta}}\,
\label{boundsHiggs}
\eeq
for $H\to \bar{f} f+G$ and $H\to VV+G$, respectively.
In the limit  $x_f^2,~x_V^2 \to 0$ and $x_\Delta=0$, we get
\beq
M_H < M_D \times \left(5.5,~5.0,~4.5,~4.2,~3.9\right),~~~
M_H < M_D \times \left(5.6,~5.0,~4.6,~4.2,~3.9\right),
\label{bounds_unH}
\eeq
for $\delta=2,3,4,5,6$, respectively.

As an application of our results, we can consider 
two representative scenarios in the SM: the 
light ($M_H < 2 M_W$) and the heavy ($M_H > 2 M_t$) Higgs boson.
In particular, we set  $M_H=120$ GeV and $M_H=500$ GeV, respectively.
Then, approximating the total $H$ width by the dominant tree-level
$\Gamma(b\bar b)$ and $\Gamma(t \bar t+WW +ZZ)$, respectively,
the gravitational decays BR's are given by
\begin{itemize}
\item {\bf Light Higgs  ($M_H< 2 M_W$)}
\beq
{\rm BR}(H\to \bar{b} b+ G)=
\left(\frac{S_{\delta-1}}{192\pi^2\rho_f(x_b)}\right)
\left(
\frac{M_H}{M_D}\right)^{2+\delta}I_H^F(x_{\Delta},x_b,\delta)
\eeq
\item {\bf Heavy Higgs  ($M_H > 2 M_t$)}
\bea
{\rm BR}(H\to WW+ G)&=&
\left(\frac{ S_{\delta-1}}{24\pi^2\Delta_H}\right)
\left(
\frac{M_H}{M_D}\right)^{2+\delta}\, I_H^V(x_{\Delta},x_W,\delta)
\\
{\rm BR}(H\to ZZ+ G)&=&
\left(\frac{ S_{\delta-1}}{48\pi^2\Delta_H}\right)
\left(
\frac{M_H}{M_D}\right)^{2+\delta}\, I_H^V(x_{\Delta},x_Z,\delta)
\\
{\rm BR}(H\to \bar{t}t+ G)&=&
\left(\frac{S_{\delta-1} x_t}{16\pi^2\Delta_H}\right)
\left(
\frac{M_H}{M_D}\right)^{2+\delta}\, I_H^f(x_{\Delta},x_t,\delta)
\eea
\end{itemize}
where the variables $x_i$ are defined as 
$x_i=M_i^2/M_H^2$, with $i=b,t,W,Z$, and 
$\Delta_H=12 x_t \rho_f(x_t)+2\rho_V(x_W)+\rho_V(x_Z)$.
Note that, in eqs.(\ref{hvv}) and (\ref{hvvdue}),
$G_V \to G_F$ for the $H$ decaying both into $W$'s and into $Z$'s.

In the case $\delta=2$, we obtain
\bea
{\rm BR}(H\to \bar{b} b+ G_X)&=&
\frac{2.2\times 10^{-7} }{M^4_D({\rm TeV})},~~~~M_H=120~{\rm GeV}
\\
{\rm BR}(H\to W~W+ G_X)&=&
\frac{2.1\times 10^{-5} }{M^4_D({\rm TeV})},~~~~M_H=500~{\rm GeV}
\\
{\rm BR}(H\to Z~Z+ G_X)&=&
\frac{8.7\times 10^{-6} }{M^4_D({\rm TeV})},~~~~M_H=500~{\rm GeV}
\\
{\rm BR}(H\to \bar{t}~t+ G_X)&=&
\frac{8.7\times 10^{-7} }{M^4_D({\rm TeV})},~~~~M_H=500~{\rm GeV}
\label{BR_H_2}
\eea
where we used,
for $M_H=120$ GeV, $I_H^f(0,x_b,2)=0.312$, and,
for $M_H=500$ GeV,
$I_H^V(0,x_W,2)=0.0393$, $I_H^V(0,x_Z,2)=0.0321$, and $I_H^f(0,x_t,2)=0.00876$.

We can see that, in order to constrain 
$M_D$ in the range of a few TeV's
for $\delta=2$, we need a sensitivity on the Higgs BR's 
of order 
${\cal O}(10^{-7})$ and ${\cal O}(10^{-5})$, for the light and heavy
Higgs boson, respectively. 
Higher sensitivities are needed to explore the case of 
a larger $\delta$.
Such sensitivities on the Higgs BR's are  beyond the reach
of any presently planned experiment by a few orders of magnitude.
Anyhow, they could become of some interest for physics 
that might be
studied at a Higgs boson factory in a not-near future 
(see, e.g.,\cite{factory}).
\section{Conclusions}
In this paper, we studied the effects of quantum gravity propagating
in large extra dimensions in a few favoured decay channels of heavy particles.
In particular, we analyzed the inclusive radiative emission of 
Kaluza-Klein spin-2 gravitons in the following decay channels:
the two-fermions decays of
massive gauge bosons, heavy quarks,   
Higgs bosons, and  the two-massive gauge bosons decay of Higgs bosons.

Due to the huge number of KK gravitons radiated, the inclusive widths,
for a particle of mass $M$,
is only suppressed by a factor of order 
$\left(\frac{M}{M_D}\right)^{2+\delta}$, versus
the usual factor $\left(\frac{M}{M_P}\right)^{2}$ arising in quantum  gravity 
in 3+1 dimensions.
If the mass of the particle is pretty 
close to the Plank mass in D-dimensions $M_D$, the quantum gravity
effects might sizeably affect the heavy particles decays. 
In scenarios where the  SM fields 
propagate in extra dimensions with a $\sim$ TeV compactification scale,
good candidates for the decaying heavy particles might be the
KK excitations of the usual SM particles.

In this framework, we provided analytical results for the
square modulus of the amplitudes, 
and numerical results for the inclusive widths.
Final-state masses have been taken into 
account, apart from the case of a heavy fermion decay, where a 
massless final fermion is assumed.
Since, experimentally, the KK gravitons are indirectly detected by measuring 
missing energy and mass in the decay process, 
we presented plots for the distributions of the widths 
versus the KK graviton mass. 
We showed that the position of the maximum for each distribution 
is quite sensitive to the number of extra dimensions.


We also discussed the validity of the present perturbative approach
for heavier masses of the decaying particles.
We showed that, when the mass of the decaying particles is a few
times $M_D$, the radiative widths exceed the corresponding tree-level widths,
breaking unitarity.
While there are not unitarity problems in the effective theory for $M< M_D$,
one should keep in mind that, for larger masses, non perturbative effects
get in general important. 

As an application of our study, we analyzed the 
decays $Z\to \bar{f} f+G$, ~~$W\to \bar{f}^{\prime} f+G$, 
~~$t\to W b+G$, ~~$H\to \bar{f} f+G$, and $H\to WW+G$, 
that can be of interest at present and future experiments.
In the case of $Z$ decays,  the present sensitivity on the 
BR($Z\to \bar{f} f+E_{miss})$ at LEP1 can push the lower 
bound on $M_D$ from this decay channel not far from the bounds
obtained at LEP2 \cite{land}.  
Similar bounds are obtained from the top quark gravitational decay, 
assuming (quite unrealistically, at present) that some day the experimental 
sensitivity on its BR  will get close to the $Z$ one at LEP1.
For the Higgs boson decay, we considered  the two representative cases
of a light ($m_H=120$ GeV) and 
heavy ($m_H=500$ GeV) Higgs. We showed that, in order to 
set lower bounds on $M_D$ of order  a few TeV's
for $\delta=2$, a sensitivity 
$\sim 10^{-7}$ and $\sim 10^{-5}$, respectively, on its 
gravitational BR's is required.
The latter sensitivities are definitely a few orders of magnitude
beyond the reach of the planned experiments for Higgs production and study.
\section*{Acknowledgements}
We would like to thank Gian Giudice, Riccardo Rattazzi, and Claudio Scrucca
for useful discussions. E.G. acknowledges the Theory Division 
of CERN for its support during part of this work.
\newpage
\section*{Appendix A1}

In this appendix we report the Feynman rules 
for gravitational interactions which are 
relevant for the processes considered in this article. In particular,
by means of eqs.(\ref{gravity}) and (\ref{EMT}), we obtain

\vspace{1cm}
\epsfbox{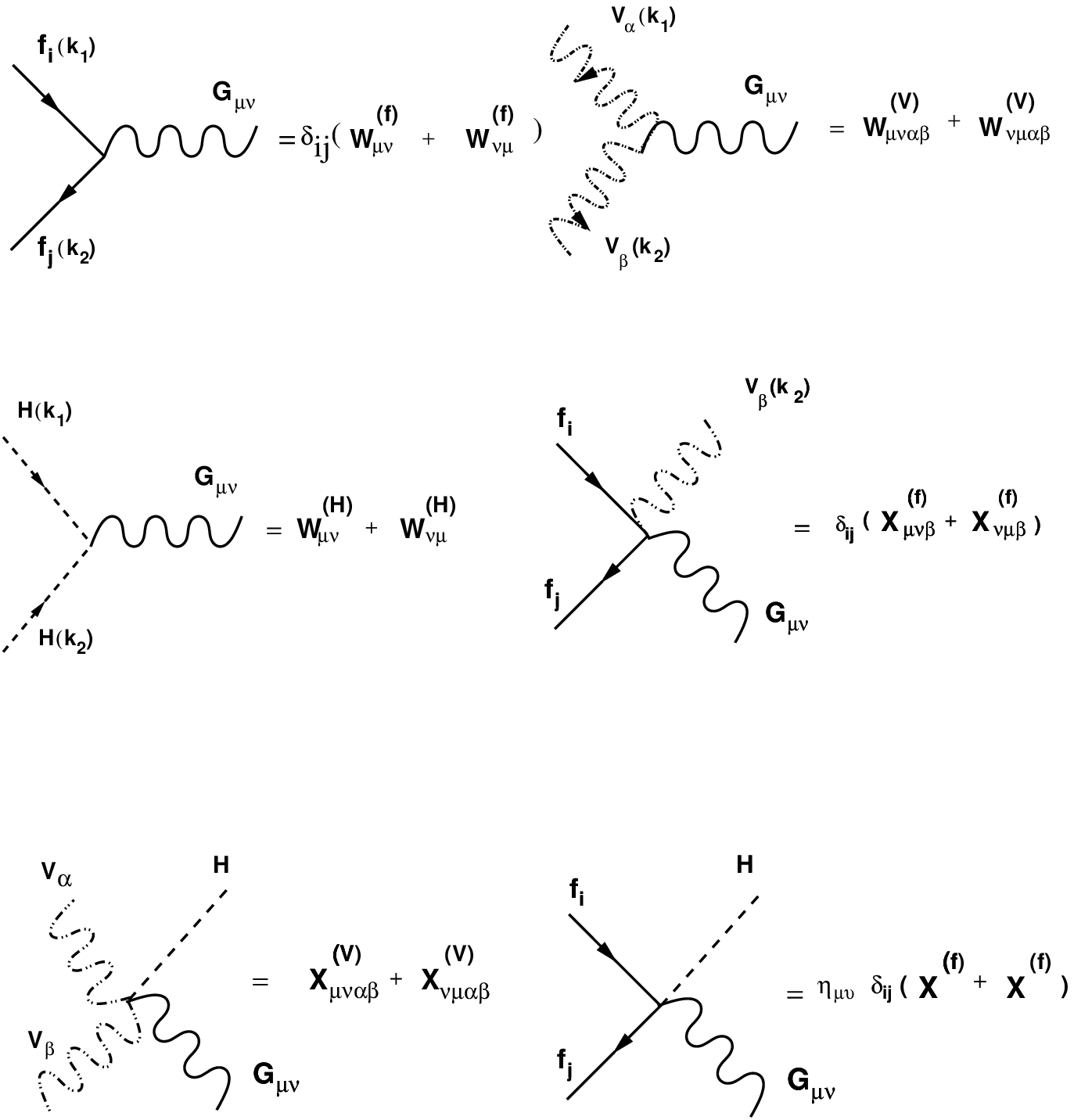}

\vspace{1cm}
\noindent
where we used the convention that 
particle momenta (indicated inside parenthesis) flow 
along the arrow directions.
The expressions of $W$ and $X$ quantities, corresponding 
to 3-line and 4-line interaction vertices,
respectively, involving 2-vectors ($W^{(V)},~X^{(V)}$), 
2-fermions ($W^{(f)},~X^{(f)}$), and 2-Higgs bosons ($W^{(H)}$), 
are given below.
\begin{itemize}
\item {\bf Vector}
\bea
W^{(V)}_{\mu\nu\alpha\beta}&=&-\frac{i}{\bar{M}_P}\left\{
\frac{1}{2}\eta_{\mu\nu}\left(k_{2\alpha}k_{1\beta} -
\eta_{\alpha\beta}\, k_1\cdot k_2\right)+
\eta_{\alpha\beta} k_{1\mu}k_{2\nu}-\eta_{\mu\beta} k_{1\nu}k_{2\alpha}
\right. \nonumber \\
&+& \left. \eta_{\mu\alpha}\left(\eta_{\nu\beta}\, k_1\cdot k_2
-k_{2\nu}k_{1\beta}\right)
+M_V^2\left(
\eta_{\mu\alpha}\eta_{\nu\beta}
-\frac{1}{2}\eta_{\mu\nu}\eta_{\alpha\beta}
\right)
\right\}
\\
X^{(V)}_{\mu\nu\alpha\beta}&=&-\frac{i}{\bar{M}_P}g\,M_V\left(
\eta_{\mu\alpha}\eta_{\nu\beta}-\frac{1}{2}\eta_{\mu\nu}\eta_{\alpha\beta}
\right)
\eea
\item  {\bf Fermion}
\bea
W^{(f)}_{\mu\nu}&=&-\frac{i}{4\bar{M}_P}\left\{
\gamma_{\mu}\left(k_{1\nu}+k_{2\nu}\right)-
\eta_{\mu\nu}\left(\hat{k}_1+\hat{k}_2-M_{f_i}\right)
\right\}
\\
X^{(f)}_{\mu\nu\alpha}&=&-\frac{i}{2\bar{M}_P}g\left(
\gamma_{\mu}\eta_{\nu\alpha} - \gamma_{\alpha}\eta_{\mu\nu}\right)
\left(g_V+g_A\gamma_5\right)
\\
X^{(f)}&=&-\frac{i}{\bar{M}_P}\lambda_{f_i}
\eea
\item  {\bf Higgs boson}
\bea
W^{(H)}_{\mu\nu}&=&-\frac{i}{\bar{M}_P}\left\{k_{1\mu}k_{2\nu}
-\frac{1}{4}\eta_{\mu\nu}\left(k_1 \cdot k_2 - M_H^2\right)
\right\}
\eea
\end{itemize}
where the symbol $\hat{p}$ stands for 
$\hat{p}\equiv \gamma^{\alpha} p_{\alpha}$.
\section*{Appendix A2}
In this appendix we report the expressions for the
functions $F_V^{(\pm)}(t,u)$, $F_f(t,u)$, and
$F_H^{V,f}(t,u)$ appearing in eqs.(\ref{Vm2}),
(\ref{Fm2}), and (\ref{Hm2})
respectively. Everywhere, the relation  $s=x_G-t-u$ holds.

\vspace{0.5cm}
\noindent
$\bullet$ ~~ ${\bf V \to \bar{f} \,\,f\,+\,G }$

\bea
F_{\rm V}^{(\pm)}(t,u)&=& \sum_{i=1}^{10}\, T^{(\pm)}_i
 \nonumber \\
 \nonumber \\
T_1^{(+)}&=&
-\frac{1}{3\, s^2}\left\{
-16 + 22\,t - 6\,{t^2} + 22\,u 
         - 2\,t\,u - 
        6\,{u^2} + 16\,x_f - 32\,t\,x_f
\right. \nonumber \\
&+& 
\left.
{t^2}\,x_f - 32\,u\,x_f + 2\,t\,u\,x_f + {u^2}\,x_f 
      +  2\,{x_G^2}\,\left( -7 + 2\,x_f \right)  
\right.\nonumber \\
&-& 
\left.
        2\,x_G\,\left( 22 - 7\,t - 7\,u - 32\,x_f + 2\,t\,x_f + 
           2\,u\,x_f \right)
\right\}
\nonumber \\
T_1^{(-)}&=&
-\frac{x_f}{3\, s^2}\left\{
48 - 56\,t + 13\,{t^2} - 56\,u + 
        26\,t\,u + 13\,{u^2} 
\right. \nonumber \\
&-& 
\left.
        4\,\left( -28 + 13\,t + 13\,u \right) \,x_G + 52\,{x_G^2}
         \right\}
\nonumber \\
T_2^{(+)}&=& 
\frac{1}{6\, t^2}\left(-1 + x_f \right)
     \left( 3\,{x_G^2} - 4\,x_G\,x_f - 32\,{x_f^2} \right)
\nonumber \\
T^{(-)}_2&=& 
\frac{x_f}{2\, t^2}
     \left( 3\,{x_G^2} - 4\,x_G\,x_f - 32\,{x_f^2} \right)
\nonumber \\
T_3^{(\pm)}&=&\left\{ T_2^{(\pm)} ~~(u \leftrightarrow t)\right\}
\nonumber \\
T_4^{(+)}&=&
 \frac{1}{6\, ts}\left\{24 + 36\,u + 12\,{u^2} + 
        {x_G^3}\,\left( 6 - 12\,x_f \right)  - 56\,x_f - 68\,u\,x_f 
\right. \nonumber \\
&-& 
\left. 
        14\,{u^2}\,x_f + 3\,{u^3}\,x_f + 32\,{x_f^2} + 
        56\,u\,{x_f^2} - 4\,{u^2}\,{x_f^2} 
\right. \nonumber \\
&-& 
\left. 
        2\,{x_G^2}\,\left( 3 + 6\,u + 2\,x_f - 12\,u\,x_f + 
           8\,{x_f^2} \right) 
\right.  \nonumber \\
&+& 
\left. 
        x_G\,\left( -12 - 6\,u + 6\,{u^2} + 48\,x_f + 18\,u\,x_f - 
           15\,{u^2}\,x_f 
\right. \right.\nonumber \\
&-& 
\left. \left.
- 112\,{x_f^2} + 16\,u\,{x_f^2} \right) 
\right\}
\nonumber \\
T_4^{(-)}&=&
- \frac{x_f}{6\, ts}\left\{
12\,{x_G^3} + 
        4\,{x_G^2}\,\left( -9 - 6\,u + 4\,x_f \right)  
\right. \nonumber \\
&+& 
\left.
        x_G\,\left( -32 + 6\,u + 15\,{u^2} + 16\,x_f - 
           16\,u\,x_f \right)
\right. \nonumber \\
&+& 
\left.
 \left( u - 6\right) \,   \left( -12 - 12\,u - 3\,{u^2} + 16\,x_f + 4\,u\,x_f
           \right)
\right\}
\nonumber \\ 
T_5^{(\pm)}&=& \left\{T_4^{(\pm)}  ~~(u \leftrightarrow t)\right\}
\nonumber \\
T_6^{(+)}&=&
\frac{1}{3\, tu }\left\{
12 + 3\,{x_G^3} - 
        18\,{x_G^2}\,\left( -1 + x_f \right)  - 60\,x_f + 
        80\,{x_f^2} - 32\,{x_f^3} 
\right. \nonumber \\
&+& 
\left.
        x_G\,\left( 27 - 75\,x_f + 32\,{x_f^2} \right)
\right\}
\nonumber \\
T_6^{(-)}&=&
-\frac{x_f}{3\, tu }\left\{
36 + 39\,x_G + 8\,{x_G^2} - 144\,x_f - 
        56\,x_G\,x_f + 96\,{x_f^2} \right\}
\nonumber \\
T_7^{(+)}&=&
\frac{1}{6\, s}\left\{
         56\,{x_G^2}\,x_f + 
        x_G\,\left( 88 - 28\,x_f - 53\,t\,x_f - 53\,u\,x_f + 
           32\,{x_f^2} \right) 
\right. \nonumber \\
&-& 
\left.
2\,\left( -42 + 10\,t - 3\,{t^2} + 10\,u + 6\,t\,u - 
           3\,{u^2} + 64\,x_f + 11\,t\,x_f 
\right. \right.  \nonumber \\
&-& 
\left. \left.
7\,{t^2}\,x_f + 
           11\,u\,x_f - 11\,t\,u\,x_f - 7\,{u^2}\,x_f - 56\,{x_f^2} + 
           6\,t\,{x_f^2} + 6\,u\,{x_f^2} \right)
\right\}
\nonumber \\
T_7^{(-)}&=&
\frac{x_f}{6\, s}\left\{
 56\,{x_G^2} + 
        x_G\,\left( 20 - 53\,t - 53\,u + 32\,x_f \right)  
\right. \nonumber \\
&-& 
\left. 
        2\,\left( 88 + 47\,t - 7\,{t^2} 
 + 47\,u - 11\,t\,u - 
           7\,{u^2} - 8\,x_f + 6\,t\,x_f + 6\,u\,x_f \right)
\right\}
\nonumber \\
T_8^{(+)}&=&
\frac{1}{6\, u}\left\{36 + 12\,t + 
        12\,{x_G^2}\,\left( -1 + x_f \right)  - 12\,x_f - 24\,t\,x_f 
\right. \nonumber \\
&+& 
\left.
        3\,{t^2}\,x_f - 14\,t\,{x_f^2} + 
        x_G\,\left( 9\,t + 30\,x_f - 9\,t\,x_f + 16\,{x_f^2} \right)
\right\}
\nonumber \\
T_8^{(-)}&=&
\frac{x_f}{6\, u}\left\{-60 - 6\,t + 3\,{t^2} + 12\,{x_G^2} - 
        160\,x_f - 14\,t\,x_f 
\right. \nonumber \\
&+& 
\left.
        x_G\,\left( -18 - 9\,t + 16\,x_f \right) \right\}
\nonumber \\
T_9^{(\pm)}&=& \left\{T_8^{(\pm)}  ~~(u \leftrightarrow t)\right\}
\nonumber \\
T_{10}^{(+)}&=& - \frac{1}{6}\left\{ 
    8 + 6\,t + 6\,u + 22\,x_f + t\,x_f + u\,x_f + 
        28\,{x_f^2} + 12\,x_G\,\left( -1 + 2\,x_f \right)
\right\}
\nonumber \\
T_{10}^{(-)}&=& - \frac{x_f}{6}\left(
130 + t + u + 24\,x_G + 28\,x_f 
\right)
\nonumber \\
\eea
$\bullet$ ~~ ${\bf f_i \to f_j \,\,V\,+\,G }$

\bea
F_{\rm f}(t,u)&=&\sum_{i=1}^{10}\, F_i
\nonumber \\
\nonumber \\
F_1&=&
\frac{1}{6\, s^2}
\left\{
{{\left(t + u - 2\,x_G \right) }^2} + 
     x_V\,\left( -12\,t + 11\,{t^2} - 9\,{u^2} + 24\,x_G - 24\,t\,x_G 
\right.\right. \nonumber \\
&+& 
\left. \left.
        4\, {x_G^2} + 2\,u\,\left( -6 + t + 8\,x_G \right)  \right) 
       - 4\,\,{x_V^2}\, \left( -4 + 8\,t + 3\,{t^2} + 3\,{u^2} 
\right.\right. \nonumber \\
&+& 
\left. \left.
        u\,\left( 8 + t - 7\,x_G \right)  - 16\,x_G - 7\,t\,x_G + 
        7\,{x_G^2} \right) 
\right. \nonumber \\
&+& 
\left.
     4\,{x_V^3}\,\left( 4 + 11\,t + 11\,u - 22\,x_G \right)  - 
     32\,{x_V^4}
\right\}
\nonumber \\
F_2 &=&
\frac{1}{12\, t^2}\left( -32 - 4\,x_G + 3\,{x_G^2} \right) \,
     \left( -1 - x_V + 2\,{x_V^2} \right)
\nonumber \\
F_3&=&
\frac{{x_G^2}}{4\, u^2}\,\left( -1 - x_V + 2\,{x_V^2} \right)
\nonumber \\
F_4&=&
\frac{1}{6\, ts}\left\{12 - 3\,{u^3} - 4\,x_G - 2\,{x_G^2} + 6\,{x_G^3} + 
     4\,{u^2}\,\left( -1 + 3\,x_G \right)  
\right. \nonumber \\
&+& 
\left.
     u\,\left( 2 + 6\,x_G - 15\,{x_G^2} \right)  - 
     2\,x_V\, \left( -10 - 32\,x_G - 8\,{x_G^2} + 3\,{x_G^3} 
\right.\right. \nonumber \\
&+& 
\left. \left.
        {u^2}\,\left( -4 + 3\,x_G \right)  + 
        u\,\left( 14 + 12\,x_G - 6\,{x_G^2} \right)  \right)  
\right. \nonumber \\
&-& 
\left.
     2\,{x_V^2}\, \left( 2 + 6\,{u^2} + 12\,x_G - 3\,{x_G^2} - 
        u\,\left( 7 + 3\,x_G \right)  \right)  
\right. \nonumber \\
&-& 
\left.
     4\,{x_V^3}\, \left( 1 + 9\,u - 3\,x_G \right)  - 24\,{x_V^4}\right\}
\nonumber \\
F_5&=&\frac{1}{us}\left\{
2 + 6\,x_G + 5\,{x_G^2} + {x_G^3} + 
  \frac{t^2}{2}\,\left( 2 + x_G \right) - 
  \frac{t}{2}\,\left( 6 + 10\,x_G + 3\,{x_G^2} \right) 
\right. \nonumber \\
&-& 
\left.
\,x_V\, \left( 2 + {t^2}\,\left( -1 + x_G \right)  + 4\,x_G + 2\,{x_G^2} + 
     {x_G^3} - t\,x_G\,\left( 1 + 2\,x_G \right)  \right) 
\right. \nonumber \\
&+& 
\left.
{x_V^2}\,  \left( -6 - 2\,{t^2} - 4\,x_G + {x_G^2} + 
     t\,\left( 9 + x_G \right)  \right)  + 
  2\,{x_V^3}\, \left( 5 - 3\,t + x_G \right)  - 4\,{x_V^4}\right\}
\nonumber \\
F_6&=&\frac{1}{2\, tu}\left\{4 + x_G - {x_G^2} - 2\,
    x_V\,\left( 2 + 4\,x_G - 3\,{x_G^2} + {x_G^3} \right) 
\right. \nonumber \\
&+& 
\left.
     {x_V^2}\, \left( -12 + 25\,x_G - 12\,{x_G^2} \right)  + 
     {x_V^3}\, \left( 20 - 18\,x_G \right)  - 8\,{x_V^4}\right\}
\nonumber \\
F_7&=&
\frac{1}{6\,s}\left\{-10\,{u^2} + u\,\left( 34 - 11\,t + 31\,x_G \right)
\right. \nonumber \\
&-& 
\left.
     2\,\left( 8 - 22\,t + 2\,{t^2} + 36\,x_G - 11\,t\,x_G + 
        14\,{x_G^2} \right)  
\right. \nonumber \\
&-& 
\left.
     2\,x_V\,\left( 6 - 5\,t + 3\,{t^2} - 
        6\, \left( 1 + t \right) \,u + 3\,{u^2} - 7\,x_G \right) 
\right. \nonumber \\
&+& 
\left.
4\,{x_V^2}\, \left( 16 + 5\,t + 5\,u - 22\,x_G \right)
       - 84\,{x_V^3}\right\}
\nonumber \\
F_8&=&\frac{1}{u}\left\{
-3 + t - \frac{9}{2}\,x_G + \frac{t}{4}\,x_G - {x_G^2} + 
 x_V\, \left( t - \frac{3}{2}\,t\,x_G + 
     \frac{x_G}{2}\,\left( 3 + 4\,x_G \right) \right) 
\right. \nonumber \\
&+& 
\left.
 {x_V^2}\, \left( 9 - 2\,t \right)  - 6\,{x_V^3}\right\}
\nonumber \\
F_9&=& 
\frac{1}{t}\left\{
\frac{1}{12}\left(-6\,{u^2} + u\,\left( 2 + 15\,x_G \right)  + 
      6\,\left( 10 + x_G - 2\,{x_G^2} \right)\right) 
\right. \nonumber \\
&-& 
\left.
  \frac{x_V}{2}\left( 3\,u\,\left( -2 + x_G \right)  + 
        \left( 13 - 4\,x_G \right) \,x_G \right)  - 
 {x_V^2}\, \left( 7 + 2\,u \right)  - 6\,{x_V^3}\right\}
\nonumber \\
F_{10}&=& 
\frac{1}{6}\left\{55 + 2\,t - u + 12\,x_G + 
    x_V\, \left( 11 + 6\,t + 6\,u - 12\,x_G \right)  + 8\,{x_V^2}\right\}
\eea
\\
$\bullet$ ~~ ${\bf H \to V\,\,V\,+\,G }$

\bea
F_{\rm H}^V(t,u)&=&\sum_{i=1}^{10}\, H^V_i 
\nonumber \\
\nonumber \\
H^V_1&=&
\frac{1}{4\, s^2}\left\{
\left( 16 + {t^2} + 20\,u + {u^2} + 
       2\,t\,\left( 10 + u \right)  - 
       4\,\left( 7 + t + u \right) \,x_G + 4\,{x_G^2} \right) 
\right. \nonumber \\
&\times& 
\left.
     \left( 1 - 4\,x_V + 12\,{x_V^2} \right)
\right\}
\nonumber \\
H^V_2&=& 
\frac{1}{t^2}\left\{
{x_G^2}\,\left( {\frac{1}{4}} - 6\,x_V + 13\,{x_V^2} \right)  + 
  {x_V^2}\,\left( 4 - 5\,{u^2} - 16\,x_V + 48\,{x_V^2} \right)  
\right. \nonumber \\
&+& 
\left.
  x_G\,x_V\,\left( 3 + 5\,u - 12\,x_V + 56\,{x_V^2} \right)
\right\}
\nonumber \\
H^V_3&=&\left\{H^V_2  ~~(u \leftrightarrow t)\right\}
\nonumber \\
H^V_4&=&
\frac{1}{ts}\left\{
3 - 16\,x_V - {u^3}\,x_V + 52\,{x_V^2} - 48\,{x_V^3} + 
  {x_G^3}\,\left( -1 + 6\,x_V \right)  
\right. \nonumber \\
&+& 
\left.
  {u^2}\,\left( 3 - 11\,x_V + 14\,{x_V^2} \right)  + 
  \frac{3\, x_G^2}{2}\,\left( 4 + u - 24\,x_V - 6\,u\,x_V + 
        32\,{x_V^2} \right) 
\right. \nonumber \\
&+&
\left. 
  u\,\left( 6 - 23\,x_V + 48\,{x_V^2} - 12\,{x_V^3} \right)  + 
  x_G\,\left( {u^2}\,\left( -{\frac{1}{2}} + 5\,x_V \right)  
\right.\right. \nonumber \\
&+&
\left. \left.
     u\,\left( -{\frac{17}{2}} + 41\,x_V - 52\,{x_V^2} \right)  + 
     4\,\left( -2 + 11\,x_V - 23\,{x_V^2} + 6\,{x_V^3} \right) 
      \right)
\right\}
\nonumber \\
H^V_5&=& \left\{H^V_4 ~~(u \leftrightarrow t)\right\}
\nonumber \\
H^V_6&=&
\frac{1}{t u }\left\{
3 - 24\,x_V + 92\,{x_V^2} - 176\,{x_V^3} + 96\,{x_V^4} + 
  {x_G^2}\,\left( {\frac{1}{2}} + 2\,x_V + 26\,{x_V^2} \right) 
\right. \nonumber \\
&+& 
\left.
  x_G\,\left( 3 - 4\,x_V - 44\,{x_V^2} + 112\,{x_V^3} \right)
\right\}
\nonumber \\
H^V_7&=&
\frac{1}{s}\left\{
24 + 19\,u + {u^2} + {x_G^2}\,\left( 5 - 22\,x_V \right)  + 
  {t^2}\,\left( 1 - 4\,x_V \right) 
\right. \nonumber \\
&-&
\left. 
94\,x_V - 69\,u\,x_V -
  4\,{u^2}\,x_V + 240\,{x_V^2} + 114\,u\,{x_V^2} - 24\,{x_V^3} 
\right. \nonumber \\
&+&
\left. 
  t\,\left( 19 + 2\,u - 69\,x_V - 6\,u\,x_V + 114\,{x_V^2} \right)
\right. \nonumber \\
&+&
\left.       
\frac{x_G}{2}\,\left( -60 - 9\,u + 252\,x_V + 34\,u\,x_V - 
        424\,{x_V^2} + t\,\left( -9 + 34\,x_V \right)  \right)
\right\}
\nonumber \\
H^V_8&=&
\frac{1}{u}\left\{
6 + {x_G^2}\,\left( 1 - 6\,x_V \right)  - 16\,x_V - 7\,{t^2}\,x_V + 
  20\,{x_V^2} + 72\,{x_V^3} 
\right. \nonumber \\
&+&
\left. 
  t\,\left( 3 - 6\,x_V + 4\,{x_V^2} \right)  + 
  x_G\,\left( -8 + 29\,x_V + 8\,{x_V^2} + 
     t\,\left( -{\frac{1}{2}} + 10\,x_V \right)  \right)
\right\}
\nonumber \\
H^V_9&=& \left\{H^V_8 ~~(u \leftrightarrow t)\right\}
\nonumber \\
H^V_{10}&=&
19 + t + u - 70\,x_V - 9\,t\,x_V - 9\,u\,x_V + 126\,{x_V^2} + 
  x_G\,\left( -3 + 22\,x_V \right)
\nonumber
\eea
\\
$\bullet$ ~~ ${\bf H \to \bar{f}\,\,f\,+\,G }$

\bea
F_{\rm H}^f(t,u)&=&\sum_{i=1}^{10}\, H^f_i 
\nonumber \\
\nonumber \\
H^f_1&=&
\frac{1}{s^2}
\left( 4 + t + u - 2\,x_G \right)^2
\nonumber \\
H^f_2&=& 
\frac{1}{t^2}\left\{
{\frac{-3\,{x_G^2}}{2}} + 2\,x_G\,x_f + 16\,{x_f^2}
\right\}
\nonumber \\
H^f_3&=&\left\{H^f_2 ~~(u \leftrightarrow t)\right\}
\nonumber \\
H^f_4&=&
\frac{1}{ts}\left\{
-2\,\left( -2 - u + 2\,x_G \right) \,
  \left( 3\,\left( 1 + u \right)  - \left( 4 + u \right) \,x_f + 
    x_G\,\left( -3 + 2\,x_f \right)  \right)
\right\}
\nonumber \\
H^f_5&=& \left\{H^f_4 ~~(u \leftrightarrow t)\right\}
\nonumber \\
H^f_6&=&
\frac{1}{t u }\left\{
12 + x_G\,\left( 9 - 8\,x_f \right)  - 48\,x_f + 32\,{x_f^2}
\right\}
\nonumber \\
H^f_7&=&
\frac{1}{s}\left\{
72 + 28\,t + {t^2} + 28\,u + 2\,t\,u + {u^2} + 4\,{x_G^2} - 
  6\,\left( 4 + t + u \right) \,x_f 
\right. \nonumber \\
&+& 
\left.
  4\,x_G\,\left( -14 - t - u + 4\,x_f \right)
\right\}
\nonumber \\
H^f_8&=&
\frac{1}{u}\left\{
6\,\left( 3 + t \right)  + \left( 16 - 7\,t \right) \,x_f + 
  x_G\,\left( -15 + {\frac{3\,t}{2}} + 8\,x_f \right)
\right\}
\nonumber \\
H^f_9&=& \left\{H^f_8 ~~(u \leftrightarrow t)\right\}
\nonumber \\
H^f_{10}&=&39 - 5\,t - 5\,u - 14\,x_f
\nonumber
\eea
\newpage

\newpage
\begin{figure}[tpb]
\makefigs4{3.1in}{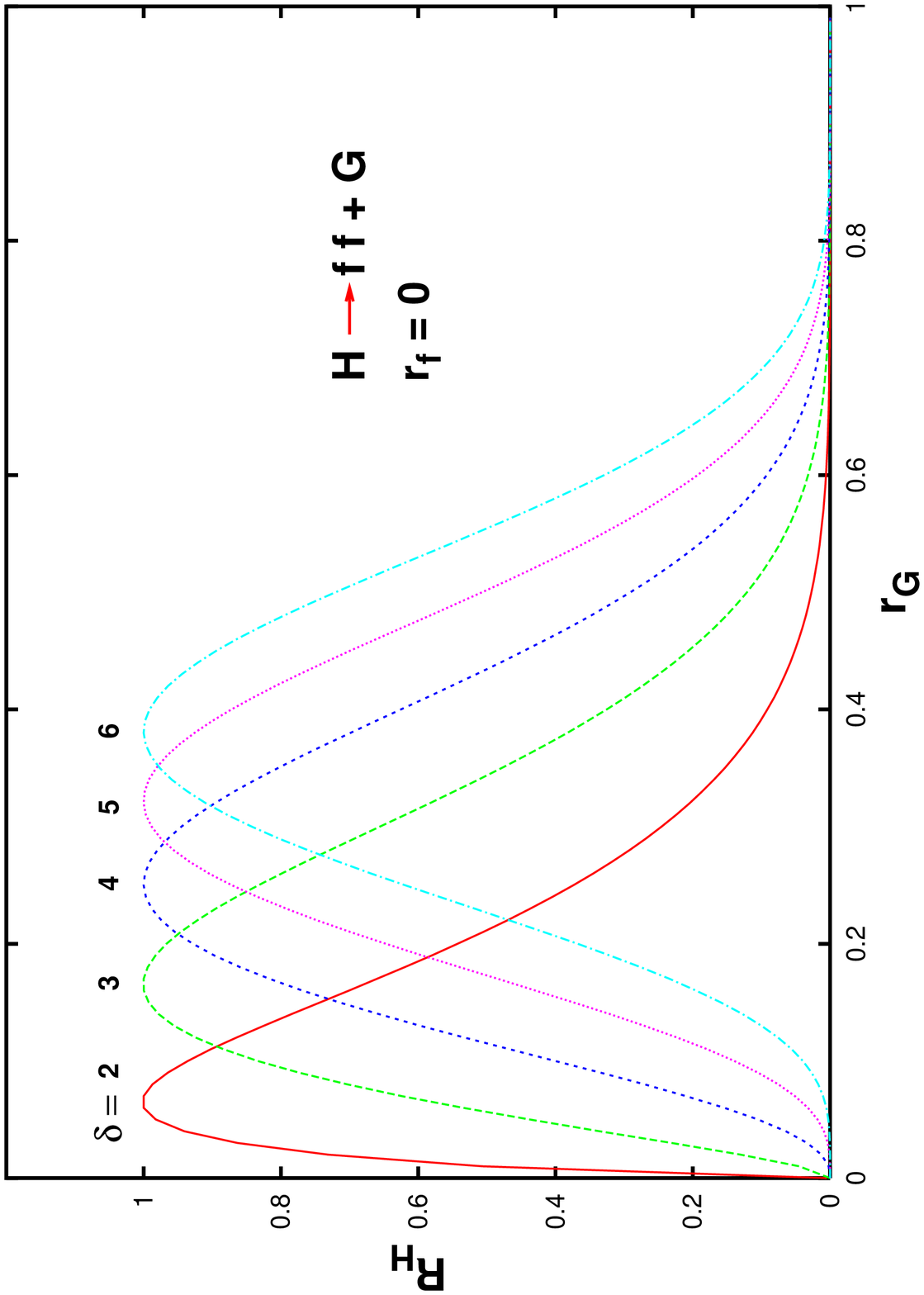 }{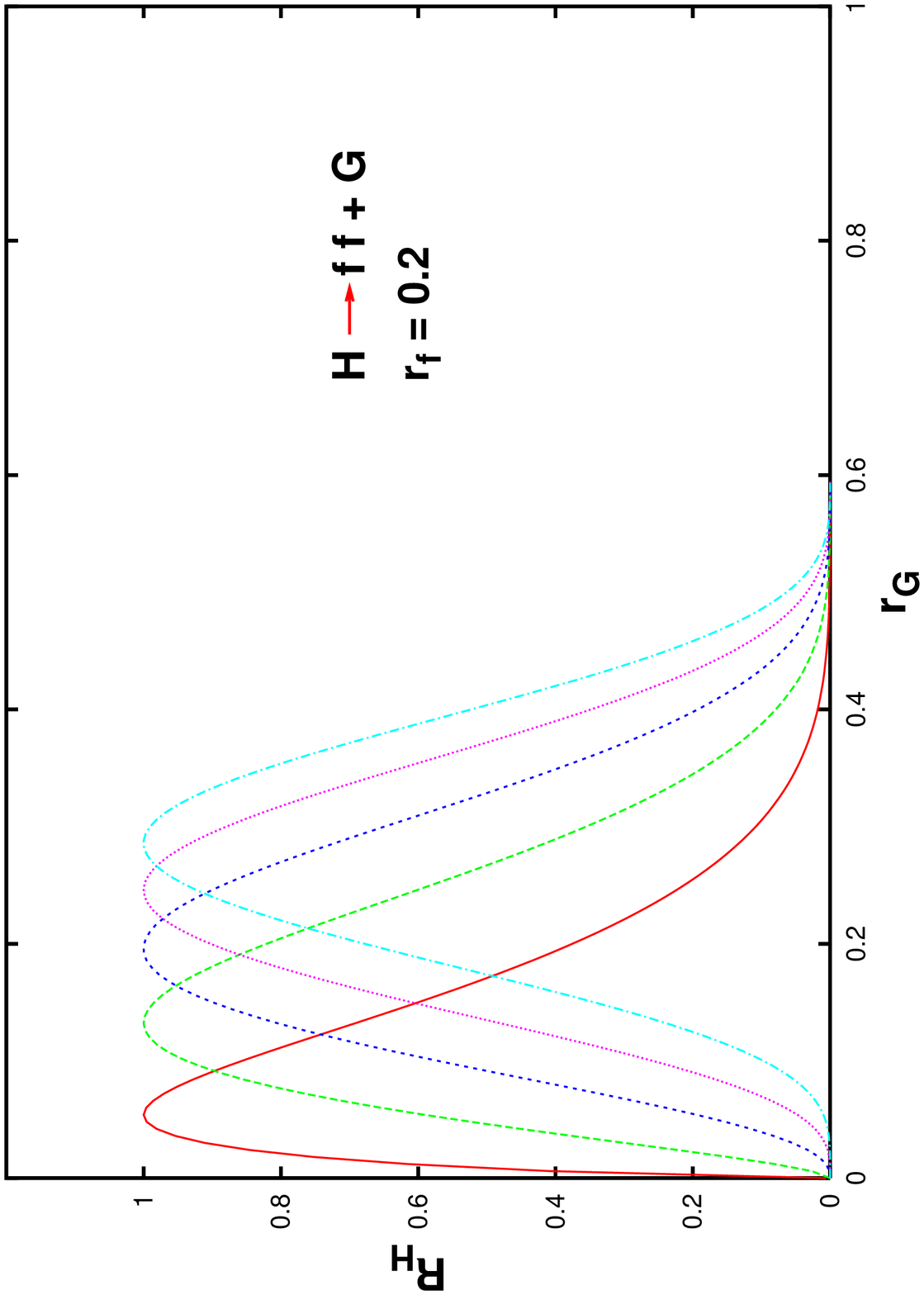 }
{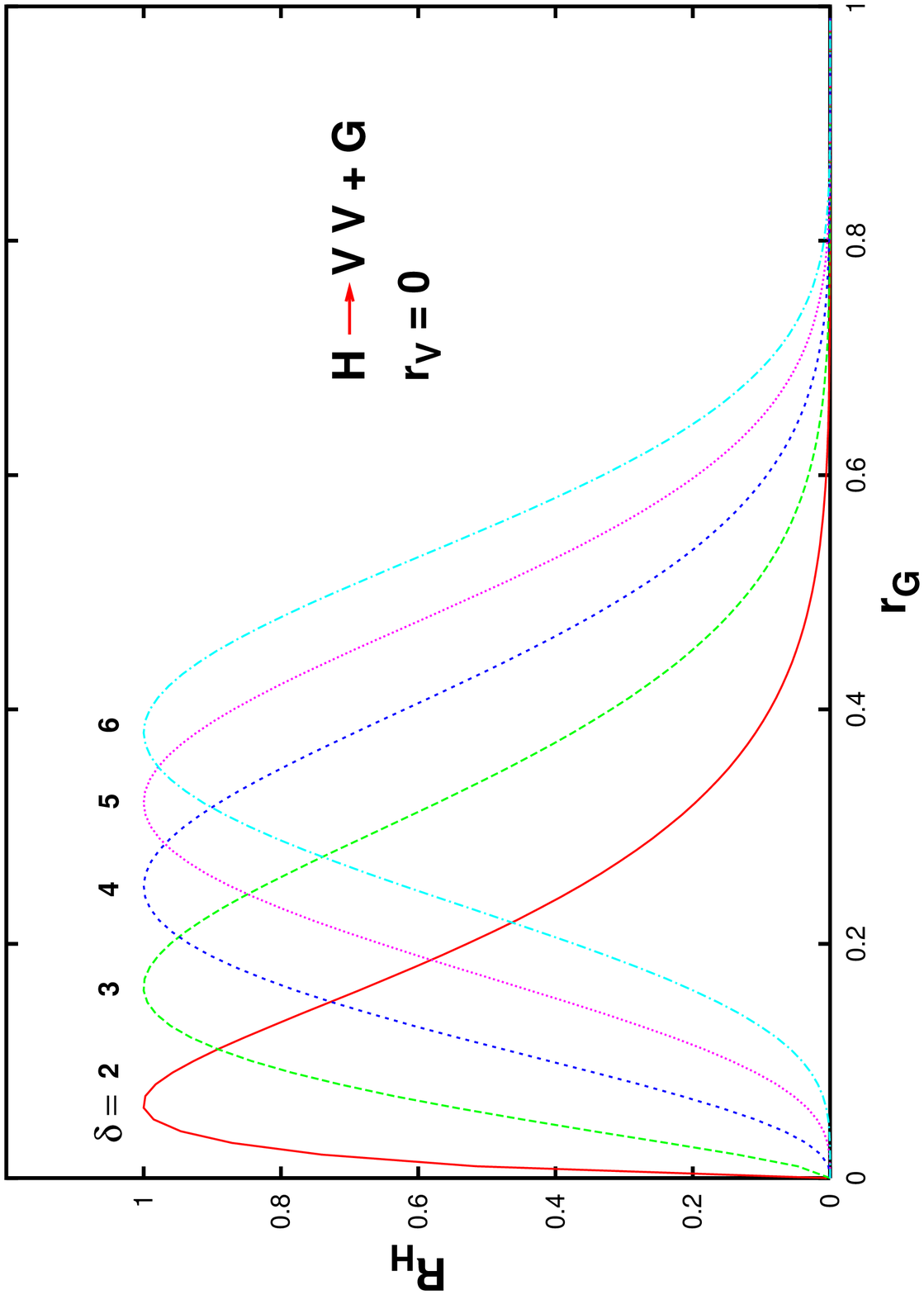 }{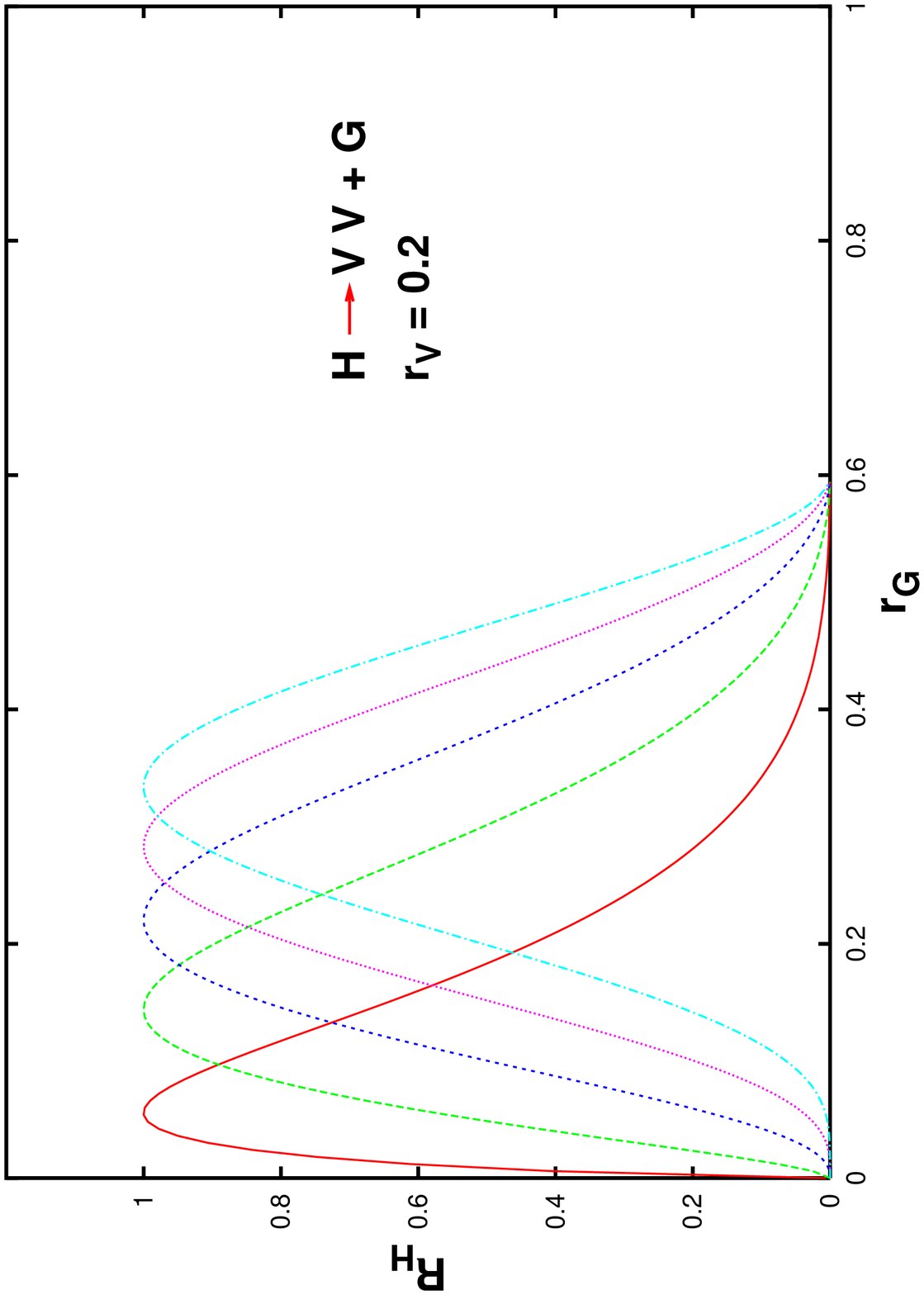 }
\caption{{\small Width distributions as in figure (\ref{fig3}), but for the
processes $H\to \bar{f} f +G$,
and $H\to VV  +G$.
}}
\label{fig4} 
\end{figure}

\begin{thebibliography}{20}

\bibitem{ADD} N. Arkani-Hamed, S. Dimopoulos, and G. Dvali, 
Phys. Lett. {\bf B429} (1998) 263.
\bibitem{GRW} G. F. Giudice, R. Rattazzi, and J. D. Wells, 
Nucl. Phys. B {\bf 544} (1999) 3.
\bibitem{HLZ} 
T. Han, J. D. Lykken, and R. Zhang, Phys. Rev. D {\bf 59} (1999) 105006.
\bibitem{PHEN_ED} E. A. Mirabelli, M. Perelstein, 
and M. E. Perskin, Phys. Rev. Lett. {\bf 82} (1999) 2236;
J. L. Hewett, Phys. Rev. Lett. {\bf 82} (1999) 4765;
E. Dudas and J. Mourad, Nucl. Phys. B {\bf 575} (2000) 3;
E. Accomando, I. Antoniadis, and K. Benakli, Nucl. Phys. B {\bf 579} (2000) 3;
S. Cullen, M. Perelstein, and M. E. Peskin, Phys. Rev. D {\bf 62} (2000)
055012; W. D. Goldberger and M. B. Wise, Phys. Lett. {\bf B475} (2000) 275; 
B. Grzadkowski and J.F. Gunion, Phys. Lett. {\bf B473} (2000) 50;
G. F. Giudice, R. Rattazzi, and J. D. Wells, hep-ph/0112161.
\bibitem{higgs} G. F. Giudice, R. Rattazzi, and J. D. Wells, 
Nucl. Phys. B {\bf 595} (2001) 250.
\bibitem{AS} For a realization of the Higgs graviscalar mixing in 
string theory see, I. Antoniadis, R. Sturani, hep-ph/0201166.
\bibitem{ED_string}
I. Antoniadis, N. Arkani-Hamed, S. Dimopoulos, and G. Dvali, 
Phys. Lett. {\bf B436} (1998) 257; 
G. Shiu and S. H. Tye, Phys. Rev. {\bf D58} (1998) 106007; 
G. Shiu, R. Shrock, and S. H. Tye, Phys. Lett.  {\bf B458} (1999) 274;
Z. Kakushadze and S. H. Tye,  Nucl. Phys. {\bf B548} (1999) 180;
I. Antoniadis, C. Bachas, and E. Dudas, Nucl. Phys. {\bf B560} (1999) 93; 
G. Aldazabal, L. E. Ibanez and F. Quevedo, JHEP {\bf 0001} (2000) 031.
\bibitem{Gravity_test} 
S. Dimopoulos and G. F. Giudice, Phys. Lett. {B379} (1996) 105
J.C. Long, H. W. Chan and J. C. Price, Nucl.Phys.B539:23-34,1999. 
\bibitem{RS} L. Randall and R. Sundrum, Phys. Rev. Lett. {\bf 83} (1999) 3370;
Phys. Rev. Lett. {\bf 83} (1999) 4690.
\bibitem{RS_phen} H. Davoudiasl, J. L. Hewett, and T. G. Rizzo
Phys. Rev. Lett. {\bf 84} (2000) 2080; Phys. Lett. {\bf B473} (2000) 
43; Phys. Rev. {\bf D63} (2001) 075004.
\bibitem{fatbrane} A. De Rujula, A. Donini, M.B. Gavela, and S. Rigolin,
Phys. Lett. {\bf B482} (2000) 195.
\bibitem{rizzo} T. G. Rizzo, Phys. Rev. {\bf D64} (2001) 095010.

\bibitem{TH_SM_ED} I. Antoniadis, Phys. Lett {\bf B246} (1990) 377;
I. Antoniadis, C. Mu\~noz, and M. Quiros, Nucl. Phys. {B397} (1993) 515;
I. Antoniadis and K. Benakli, Phys. Lett. {B326} (1994) 69; 
I. Antoniadis, K. Benakli, and M. Quiros, Phys. Lett. {B331} (1994) 313;
K. Benakli Phys. Lett. {B386} (1996) 106;
T. Appelquist, H.-C. Cheng, and B. A. Dobrescu,
Phys. Rev. {\bf D64} (2001) 035002; R. Barbieri, L.J. Hall, and 
Y. Nomura, Phys. Rev. {\bf D63} (2001) 105007;
\bibitem{DDG1} K. R. Dienes, E. Dudas, T. Gherghetta,
Nucl. Phys. {\bf B537} (1999) 47.
\bibitem{DDG2} K. R. Dienes, E. Dudas, T. Gherghetta,
Nucl. Phys. {\bf B557} (1999) 25.
\bibitem{PHEN_SM_KK}
P. Nath and M. Yamaguchi, Phys. Rev. {\bf D60} (1999) 116006; 
M. Masip and A. Pomarol, Phys. Rev. {\bf D60} (1999) 096005; 
W. J. Marciano, Phys. Rev. {\bf D60} (1999) 093006;
L. Hall and C. Kolda, Phys. Lett. {\bf B459} (1999) 213;
R. Casalbuoni, S. DeCurtis, and D. Dominici, Phys. Lett. 
{\bf B462} (1999) 48; A. Strumia, Phys. Lett. {\bf B466} (1999) 107;
F. Cornet, M. Relano, and J. Rico, Phys. Rev. {\bf D61} (2000) 037701; 
C. D. Carone, Phys. Rev. {\bf D61} (2000) 015008; 
A. Delgado, A. Pomarol, and M. Quiros, JHEP {\bf 0001} (2000) 030;
T. G. Rizzo and J. D. Wells, Phys. Rev. {\bf D61} (2000) 016007.
\bibitem{Zpole} C. Bal\'azs, H.-J. He, W. W. Repko, C.-P. Yuan,
Phys. Rev. Lett. {\bf 83} (1999) 2112.

\bibitem{veltman} H. van Dam and M. Veltman, Nucl. Phys. {\bf B22} (1970) 397.
\bibitem{form}
J.A.M. Vermaseren, {\it Symbolic Manipulation
with FORM}, published by CAN (Computer Algebra Nederland), Kruislaan
413, 1098 SJ Amsterdam, 1991, ISBN 90-74116-01-9.
\bibitem{kobel}
M. Kobel, arXiv:hep-ex/9611015, and references therein.
\bibitem{land}
G. Landsberg, arXiv:hep-ex/0105039, and references therein.

\bibitem{Beneke:2000hk}
M.~Beneke {\it et al.},
Proc. of the Workshop on Standard Model Physics (and more) at the LHC,
G.Altarelli and M.L.Mangano editors,
report CERN 2000-004, p. 419-529,
arXiv:hep-ph/0003033.

\bibitem{factory}
``Higgs Factory 2001 Snowmass  Report",\\
http://www.physics.ucla.edu/hep/hfactory/index.html
\end{thebibliography}
\end{document}